\documentclass[%
reprint,
 amsmath,amssymb,
 aps,
]{revtex4-1}

\usepackage{amssymb}
\usepackage{graphicx}
\usepackage{dcolumn}
\usepackage{bm}
\usepackage{adjustbox}
\usepackage{mathtools,tikz}
\usepackage{physics}
\usepackage{gensymb}
\newcommand{\tabitem}{~~\llap{\textbullet}~~}
\usepackage{makecell}
\usepackage[unicode=true,colorlinks=true,linkcolor=blue,citecolor=blue, urlcolor= blue, pdfencoding=auto, psdextra]{hyperref}

\begin{document}

\title{Multi-scale modeling of the effects of temperature, radiation flux and sink strength on point-defect and solute redistribution in dilute Fe-based alloys}

\author{Liangzhao Huang}
 \email{liangzhao.huang@cea.fr}
\author{Maylise Nastar}
\author{Thomas Schuler}
\affiliation{
  Universit\'e Paris-Saclay, CEA, Service de Recherches de M\'etallurgie Physique, 91191, Gif-sur-Yvette, France}
\author{Luca Messina}
\affiliation{
  CEA, DEs, IRESNE, DEC-Service d'\'Etudes et de Simulation du Comportement des Combustibles, Cadarache F-13108 Saint-Paul-Lez-Durance, France}

\begin{abstract}
In this work, we investigate the radiation-induced segregation (RIS) resulting from the coupling between the atomic and point defect (PD) fluxes towards the structural defects of the microstructure. This flux coupling depends on the migration mechanisms of PDs and atoms, including thermal diffusion mechanisms and forced atomic relocations (FAR) occurring in displacement cascades.
We derive an analytic model of the PD and solute RIS profiles accounting for PD production and mutual recombination, the FAR mechanism, and the overall sink strength of the microstructure controlling the elimination of PDs at structural defects. 
From this model, we present a parametric investigation of diffusion and RIS properties in dilute Fe-$B$ ($B$ = P, Mn, Cr, Si, Ni, and Cu) binary alloys, in the form of quantitative temperature/radiation flux/sink strength maps. As in previous works, we distinguish three kinetic domains for the diffusion and RIS properties: the recombination domain, the sink domain, and the thermal domain. Both our analytical approach and numerical applications demonstrate that the diffusion and RIS behaviors of PDs and solute atoms largely differ from one kinetic domain to another. 
Moreover, at high radiation flux, low temperature, and large sink strength, FARs tend to destroy the solute RIS profiles and therefore reduce the overall amount of RIS by forcing the mixing of solute and host atoms, especially close to PD sinks. 
Finally, we provide quantitative criteria to emulate in-reactor RIS behaviors by ion irradiation.

\end{abstract}

\maketitle

\section{Introduction}

The radiation-induced redistribution of solute atoms in materials is largely controlled by the kinetic coupling between  fluxes of lattice point defects (PDs) and atomic fluxes~\cite{RUSSELL1984229}. PDs are created by irradiation in the form of Frenkel pairs consisting of a vacancy and a self-interstitial atom (SIA). They diffuse and interact with atoms and other PDs, as well as with the microstructure of the material~\cite{Was2007}. Irradiation therefore enhances and induces redistribution of solute atoms, and affects their interplay with the  microstructure, leading to strong modifications of the mechanical, corrosion and dimensional properties of materials~\cite{Was2007}. Long-range diffusion of atoms under irradiation is mediated by successive exchanges of atoms with nearest neighbour PDs as well as forced atomic relocation (FAR) events taking place in a displacement cascade~\cite{Martin1996,Roussel2002,Was2007,Nordlund2018,Huang2019}. The relative importance of each diffusion mechanism depends on the nature of the irradiation particles and on the rate of particle irradiation. The number of Frenkel pairs created (or displacement per atom per second in dpa/s) and the number of FAR events (or replacement per atom per second in rpa/s) are proportional to the radiation flux, which enables comparison between various  irradiation particles~\cite{Norgett1975,Martin1984,Nordlund2018-2}. 
 
Under a sustained radiation flux, the steady-state creation and elimination of PDs generates net fluxes of PDs towards the extended structural defects acting as PD sinks such as grain-boundaries (GBs), dislocation lines, dislocation loops and voids. Net fluxes of PDs make solute atoms diffuse towards or away from the PDs sinks. Since the PD-solute flux coupling differs from one chemical species to another, a change of the alloy composition occurs close to the PDs sinks. This is the so-called radiation induced segregation (RIS) phenomenon~\cite{Anthony1968,Okamoto1974,Nastar2012,Ardell2016}. Recently, we have shown that when the FAR frequency is close to the thermal PD jump frequency, FAR may either enhance or reduce the PD-solute flux coupling~\cite{Huang2019}. Note that both chemical species and PDs form a RIS profile at sinks. The RIS of PDs is systematically negative, with a concentration profile dropping to thermal equilibrium concentration at sinks. RIS occurs at every PDs sink even at very small radiation dose~\cite{DONG2015692}. Therefore, RIS is often a precursor for heterogeneous precipitation of secondary phases at PDs sinks, as for example the precipitation of the ordered phase Ni$_3$-Si in austenitic steels~\cite{Okamoto1974,Barbu1975}, and the formation of Mn-Ni-Si-rich clusters in RPV ferritic steels~\cite{Miller2007,Nishiyama2008,Lambrecht2009,Miller2013,Styman2015}. RIS can induce failure of materials through various mechanisms~\cite{Was2007,Zinkle2013}, for instance the lowering of corrosion resistance due to depletion of chromium at GBs in austenitic steels~\cite{Bruemmer1999}, material embrittlement resulting from phosphorus enrichment at GBs~\cite{Miller2007}, or the shift of the ductile-to-brittle transition temperature in reactor pressure vessel steels due to the formation of solute-rich clusters~\cite{Miller2013}. 

The sign of solute RIS, positive for solute enrichment and negative for solute depletion, is directly related to the relative magnitude of solute-vacancy and solute-SIA flux coupling~\cite{Messina2016}.  Calculation methods of flux coupling coefficients rely on the Onsager formulation of solute and PDs fluxes within the framework of the thermodynamics of irreversible processes~\cite{Onsager1931-1} where these fluxes are expressed as linear combinations of chemical potential gradients. The coefficient of proportionality between the flux of species $i$ and the gradient of the chemical potential of species $j$ is the so-called phenomenological transport coefficient denoted $L_{ij}$. These transport coefficients are material specific and depend on the diffusion mechanism~\cite{Nastar2012}. For a given system, the experimental measurement of all the transport coefficients is challenging and in most cases impossible. For instance, one cannot measure equilibrium diffusion properties mediated by SIA in metals because the equilibrium concentration of SIA is too small. Furthermore, it is almost impossible to measure the off-diagonal phenomenological coefficients that are responsible for positive vacancy-solute flux coupling because these coefficients control the solute flux only when the alloying driving force is weak, which is not the case in standard thermal diffusion experiments~\cite{Nastar2012}.  The recent progress of first-principles methods allows to compute these phenomenological transport coefficients from \textit{ab initio}~\cite{wu_high-throughput_2016,PhysRevB.95.174102,Schuler2017,Messina2014,Messina2016} energetics combined with statistical models of diffusion on a lattice. The $L_{ij}$ essentially depend on the PDs diffusion mechanism and the variation of PD jump frequencies with the local alloy composition. At steady state, the RIS factor relating the solute local concentration gradients to the local PD concentration gradient normalized by the local PD concentration is essentially a function of the phenomenological coefficients $L_{ij}$, the concentration derivatives of chemical potentials, and the solute and PD local concentrations~\cite{Nastar2012,Ardell2016}. When the RIS factor is assumed to be a constant, the amplitude of the solute concentration gradient is proportional to the normalized PD concentration gradients~\cite{Martinez2018}. Therefore, the amplitude and shape of the stationary RIS profile depends not only on the RIS factor, but also on the local concentration of PDs~\cite{Nastar2012,Martinez2018}. The evolution of the PD concentration fields depends on their mobility, the radiation flux, their mutual interaction and their interaction with the microstructure and the solutes. Among PD reactions, let us mention the mutual recombination of vacancy and SIA, the clustering of PDs leading to the formation of dislocation loops and voids, and the elimination of PDs at sinks. The analysis of PD-microstructure interactions may be simplified by introducing an effective PD sink strength governing the average PD elimination rate at all PD sinks. However, the microstructure is in constant evolution due to PD clustering, production, elimination at sinks, and their interplay with solute reactions. It is therefore crucial to take the latter phenomena into account, but up to now, there is no modeling method able to account for the evolution of both the sink microstructure and the solute redistribution. Most of the RIS models either work at fixed concentrations of PDs~\cite{Marwick1978,Wiedersich1979,English1990}, or for the most advanced ones at a fixed value of the overall PD sink strength~\cite{Wolfer1983,Grandjean1994,Martinez2018}. 
 
There are experimental studies investigating the dependence of RIS on the microstructure of the irradiated sample and the irradiation conditions, including the nature of the irradiation particles~\cite{Was2002,Jiao2018}, the radiation dose and dose rate~\cite{Rehn1979,Allen2008}, and temperature~\cite{Rehn1978,Okamoto1979}. However, it is still very difficult to obtain an accurate estimation of the PD sink strength from the observation of the microstructure due to the limitations of resolution, even for nanoscale experimental techniques. In order to obtain an accurate estimation of the sink strength, experimental measurements need to be complemented with modeling~\cite{Meslin2008}. 
Predicting the evolution of RIS in nuclear power plant materials from a direct observation of neutron irradiated materials is difficult, mainly because neutron irradiation activates the sample and the radiation exposure times of several years needed to reach a few dpas are rarely available~\cite{Was2007}. Radiation fluxes of electrons and heavy ions can be high, which allows radiation doses to reach up to hundreds of dpas in a much shorter time.
However, most of the phenomena occurring under irradiation are sensitive to the radiation flux. According to simple mean-field rate theories, the PD concentrations obtained at a low radiation flux and a given temperature are identical to the ones obtained at a higher flux provided the temperature is increased by a specific amount, which suggests that a difference in radiation flux can be compensated by a temperature shift~\cite{SIZMANN1978386,Was2007,Mansur1993}. This theory has been first applied to investigate the swelling phenomena, but it relies on the assumption that solute atoms do not interfere with the kinetics of PDs and the overall PD sink strength is fixed by the initial microstructure. 
According to this theory, there are three kinetic domains: (i) at low temperature and high radiation flux, the recombination domain in which the PD concentration is controlled by the PD recombination reaction, (ii) at intermediate temperature and low radiation flux, the sink domain in which the PD concentration is controlled by the elimination of PDs at sinks, and (iii) at high temperature and low radiation flux, the thermal domain in which the PD concentration are close to thermal equilibrium concentrations~\cite{SIZMANN1978386}.
Estimations of the temperature shift required to compensate for a large radiation flux depend on the kinetic domain of the experiment and whether the system is at steady state or in a transient state. These temperature shifts require the definition of an invariant quantity, either the bulk concentration of PDs at steady state~\cite{Was2007} or the amount of PDs absorbed by sinks~\cite{Mansur1993}. 
Attempts have been made to apply Mansur's invariant PD-absorption relation to the study of solute RIS~\cite{Was2007,Was2002,Jiao2018}. The estimation of the temperature shift was good enough to yield similar RIS profiles of Cr and Ni in 304L stainless steels, respectively irradiated with neutrons and self ions~\cite{Jiao2018}. Nevertheless, in the same publication, the authors observe that the temperature shift predicted by Mansur's invariant relation is not accurate for alloys with a high dislocation density.
Yet, a material with an initial high dislocation density seems to be more appropriate to test Mansur's invariant relation, because the high PD sink strength of a microstructure full of dislocations  is less sensitive to the radiation flux and dose, and can be considered to be fixed as assumed in Mansur's theory. A recent analytical model of steady-state RIS in the sink domain precisely predicts that solute RIS does not depend on the radiation flux, whereas PD concentration does~\cite{Martinez2018}. However, as explained by the authors, we should not ignore that an increase of the dislocation density may induce a transition from the recombination domain to the sink domain, hence, shift the system from a radiation-flux dependence to another. Therefore, there is a need for a PD-RIS model accounting for both the transitions between the various PDs kinetic domains, and the effect of the irradiation conditions and the microstructure on the RIS profile within each PDs kinetic domain.


{As an important step towards a fully consistent model of solute redistribution coupled with sink-strength evolution, we derive a novel analytical RIS model, aimed at (i) taking into account all PD reactions, solute-PD interactions, and FAR mechanisms; (ii) quantitatively studying the effect of a variation of either the sink strength, the radiation flux, or the temperature on the RIS properties; and (iii) understanding and quantifying the flux-temperature effect in experiments for each kinetic domain (recombination/sink/thermal).} 

To this end, we extend the analytical approach of Ref.\,\cite{Martinez2018} to the whole temperature/radiation flux domain by including the effect of FAR and PD recombination reactions. Furthermore, we account for the variation of the RIS factor with PD and solute concentration along the segregation profiles.
Recent developments of the self-consistent mean-field theory~\cite{Nastar2000,Nastar2005} provide a procedure to treat the interplay between thermal PD diffusion mechanisms that satisfy the microscopic detailed balance, and the FAR diffusion mechanism that satisfies the global detailed balance only. The resulting fluxes under a steady-state gradient of chemical potential are still linear combinations of gradients of chemical potentials even though the transport coefficients do no longer obey the Onsager reciprocal relations~\cite{Huang2019}. Relying on the implementation of these theoretical developments in the KineCluE code~\cite{Schuler2020}, and on the recently published DFT database of vacancy and SIA hop frequencies, we present a quantitative study of flux coupling in dilute Fe-$B$ ($B$ = P, Mn, Cr, Si, Ni, and Cu) binary alloys~\cite{Messina2019} with respect to radiation flux, temperature, and PD sink strength. Combining flux-coupling factors with the analytical RIS model leads to quantitative maps of RIS with respect to these parameters. Based on the analytical PD-RIS model and its application to the Fe alloys, we discuss the validity and relevance of the temperature-shift criteria in the three PDs kinetic domains.

This paper is organized as follows: Sec.\,\ref{Sec:Models} is devoted to
a short presentation of the methods used to compute flux-coupling coefficients and bulk concentrations of PDs at steady state, and to the derivation of our RIS analytical model. Results and discussion on diffusion and RIS properties of Fe alloys are found in Sec.\,\ref{Sec:Results_and_discussions}. A summary, concluding remarks and perspectives are given in Sec.\,\ref{Sec:Conclusions}.

\section{Models} \label{Sec:Models}
A solute RIS profile is a complex function of the local concentrations of PDs and solute atoms, the local concentration dependent diffusion coefficients, the radiation flux, and the PD sink strength. We will present the procedure that we used to compute the diffusion coefficients, the flux coupling coefficients (also called flux coupling ratios), and the RIS factor from the Onsager formulation of fluxes in a binary dilute alloy A(B).

\subsection{Diffusion properties}
Following the Onsager formulation, we write the fluxes of PDs and atoms as functions of the phenomenological transport coefficients and the chemical potential gradients. Then, we briefly introduce the cluster expansion of the transport coefficients. This expansion provides an explicit variation of the transport coefficients with respect to the local PD and solute concentrations and their thermodynamic interactions. From these fluxes, we introduce the flux coupling coefficient that relates the solute flux to the PD flux in the presence of a PD chemical potential gradient.  
In the infinitely dilute limit---when the interactions between solute atoms, PDs, and solute-PD clusters larger than pairs are ignored---we express the chemical potential gradients in terms of concentration gradients and to obtain the expressions of partial, PD, and solute diffusion coefficients.

\subsubsection{Atomic fluxes and phenomenological coefficients}
Following Onsager's formalism~\cite{Onsager1931-1,Onsager1931-2}, we express the flux $\bm{J}_\alpha$ of species $\alpha$ as a linear combination of chemical potential gradients (e.g., $\grad\mu_\beta$ for species $\beta$). We assume that fluxes arising from the vacancy (V) diffusion mechanism and from the self-interstitial atom (SIA or $\text{I}$) diffusion mechanism are additive. In a binary alloy A(B), the flux of atomic species $\alpha$ ($\alpha=A\,\,\text{or}\,\,B$), reads
\begin{equation}\label{eq:J}
    \bm{J}_\alpha=\bm{J}_\alpha^{\text{V}} + \bm{J}_\alpha^{\text{I}},
\end{equation}
with 
{
\begin{align}
    \label{eq:J_alpha_V}
    \bm{J}_\alpha^{\text{V}} &= -\frac{1}{k_\text{B} T} \sum_{\beta=A,B,V} L_{\alpha \beta}^{\text{V}}{\grad\mu_{\beta}},\\
    \label{eq:J_alpha_I}
    \bm{J}_\alpha^{\text{I}} &= -\frac{1}{k_\text{B} T} \sum_{\beta=A,B,I} L_{\alpha \beta}^{\text{I}}{\grad\mu_{\beta}}.
\end{align}
Similarly, the fluxes of vacancies and SIAs read
\begin{align}
    \label{eq:J_V}
    \bm{J}_\text{V} &= -\frac{1}{k_\text{B} T} \sum_{\beta=A,B,V} L_{\text{V} \beta}^{\text{V}}{\grad\mu_{\beta}},\\
    \label{eq:J_I}
    \bm{J}_\text{I} &= -\frac{1}{k_\text{B} T} \sum_{\beta=A,B,I} L_{\text{I} \beta}^{\text{I}}{\grad\mu_{\beta}}.
\end{align}
In Eqs.\,\eqref{eq:J_alpha_V}, \eqref{eq:J_alpha_I}, \eqref{eq:J_V}, and \eqref{eq:J_I}, the coefficients $L_{\alpha \beta}^{\text{V}}$ and $L_{\alpha \beta}^{\text{I}}$ are the phenomenological transport coefficients that characterize the diffusion mediated respectively by vacancies and SIAs. For the sake of simplicity, $L_{\alpha \text{V}}^{\text{V}}$ and $L_{\alpha \text{I}}^{\text{I}}$ are respectively denoted $L_{\alpha \text{V}}$ and $L_{\alpha \text{I}}$.
}

By using the self-consistent mean-field theory, we compute the transport coefficients from the atomic jump frequencies~\cite{Nastar2000,Nastar2005}. This theory has been applied to quantitative studies of vacancy-mediated diffusion properties~\cite{Garnier2013,Messina2014,Messina2016,Schuler2017,Schuler2020-2}, combined with the direct interstitial migration mechanism~\cite{Schuler2016,Schuler2020-2}, and the SIA diffusion mechanism in dilute~\cite{Barbe2007,Messina2019} and concentrated alloys~\cite{PhysRevB.76.054206}.  It has also been extended to diffusion mechanisms that do not satisfy the microscopic detailed balance such as the FAR mechanism~\cite{Huang2019}. The recent development of a cluster formulation of the self-consistent mean field theory and its implementation into the  KineCluE code~\cite{Schuler2020}, allows for systematic and sensitivity studies of the solute concentration, strain, and temperature effects on transport coefficients in multi-component alloys~\cite{PhysRevB.95.174102,Messina2019,Schuler2020-2}. 

Now we introduce the vacancy and SIA diffusion mechanisms, as well as the FAR diffusion mechanism. According to the  parametric study of the effect of FAR on transport coefficients, varying the range of FAR does not fundamentally change the diffusion properties, in particular the flux coupling coefficients~\cite{Huang2019}. Therefore, for the sake of simplicity, we restrict the FAR mechanism to forced exchanges of atoms with their first nearest neighbours (1NN) that can be either an atom of a different chemical species or a vacancy. Note that we ignore the FAR events with SIAs because FAR frequencies are always negligible compared to SIA thermal jump frequencies. For each Frenkel pair created, the number of FAR events, $n_\text{FAR}$, ranges from a few units for electron irradiation to several hundreds for neutron irradiation~\cite{Martin1996}. The contribution of the FAR mechanism to the atomic transport increases with $n_\text{FAR}$. In the following, we set $n_\text{FAR}=100$ unless specified otherwise.

In a dilute binary alloy A(B), we consider five different cluster configurations. The B-$d$ pair configurations correspond to a single PD ($d=\text{V}$ or $\text{I}$) bound to a single solute atom B up to a distance lower than a kinetic radius $R_k$. Whenever the distance between B and $d$ is larger than $R_k$, we consider B and $d$ as isolated monomers. Therefore, the five cluster contributions considered here are the three monomers B, V and I and the two pairs B-V and B-I. We use the code KineCluE~\cite{Schuler2020} to compute the transport coefficients of each cluster from the \textit{ab initio} atomic jump frequencies. Following the kinetic cluster expansion formulation of the transport coefficients in dilute alloys, we deduce the overall transport coefficients from the cluster transport coefficients~\cite{Schuler2020}
\begin{align}
    &L_\text{BB}^{d} = L_\text{BB}^{d,\text{pair}}C^{\text{pair}}_{\text{B}d} + L_\text{BB}^{\text{mono}}C_\text{B}^{\text{mono}},\nonumber\\
    &L_{dd} = L_{dd}^{\text{pair}}C^{\text{pair}}_{\text{B}d} + L_{dd}^{\text{mono}}C_d^{\text{mono}},\nonumber\\
    &L_{\text{B}d} = L_{\text{B}d}^{\text{pair}}C^{\text{pair}}_{\text{B}d},\nonumber\\
    &L_{d\text{B}} = L_{d\text{B}}^{\text{pair}}C^{\text{pair}}_{\text{B}d}. \label{eq:L_total_cluster}
\end{align}
{Note that cluster transport coefficients are intrinsic cluster properties and independent of the local atomic fraction $C_\alpha$ of species $\alpha=\{A,B,d\}$. The cluster atomic fractions $C^{\text{pair}}_{\text{B}d}$, $C_d^{\text{mono}}$, and $C_\text{B}^{\text{mono}}$ are deduced from $C_\alpha$, and are computed in the framework of low-temperature expansions~\cite{Ducastelle1993,Schuler2017-2, Schuler2018}.
In most irradiation conditions of interest, $C_d \ll C_\text{B}$. Therefore, the cluster atomic fractions are given by~\cite{Messina2019}:
\begin{align}
    &C^{\text{pair}}_{\text{B}d} = \frac{C_\text{B} C_d Z_{\text{B}d}}{Z_d + C_\text{B}(Z_{\text{B}d}-Z_{\text{B}d}^0)}, \nonumber\\
    &C_d^{\text{mono}} = C_d\left[ 1-\frac{C_\text{B} Z_{\text{B}d}}{Z_d + C_\text{B}(Z_{\text{B}d}-Z_{\text{B}d}^0)} \right], \nonumber\\
    \label{eq:C_cluster}&C_\text{B}^{\text{mono}} = C_\text{B},
\end{align}
where $Z_{\text{B}d}$ is the partition function of the pair B-$d$, $Z_{\text{B}d}^0$ is the number of pair configurations, $Z_\text{V}=1$ for vacancies, and $Z_\text{I}=6$ for $\langle110\rangle$-dumbbells. In this case, $L_{dd}$, $L_{\text{B}d}$ and $L_{d\text{B}}$ are proportional to $C_d$ whereas $L_\text{BB}^{d}$ can be decomposed into two parts, the first being proportional to $C_d$ and the other one independent of $C_d$. Note that the coefficient $L_\text{BB}^{\text{mono}}$ is zero unless the FAR mechanism in included, because under equilibrium conditions a substitutional solute requires the presence of PDs to diffuse. }

\subsubsection{Flux-coupling coefficients}
To investigate the flux coupling driven by an excess of vacancy or SIA, we consider the ratio between fluxes of V (respectively SIA) and solute atoms (B) in a binary alloy A(B). Before irradiation, solute atoms are mostly at local equilibrium in the vicinity of PD sinks, i.e., their gradient of chemical potential is zero. From its very beginning, irradiation produces an excess of PDs which is increases as we move away from PD sinks, leading to a gradient of PD chemical potential. Therefore, $\grad{\mu_\text{V}}$ (respectively $\grad\mu_\text{I}$) are the main diffusion driving forces, at least at the beginning of irradiation. In order to investigate the flux coupling induced by these PD driving forces, we set to zero all the other gradients of chemical potential. Hence, $J_\text{V}/J_\text{B}$ and $J_\text{I}/J_\text{B}$ are respectively given by the flux coupling coefficients~\cite{Anthony1969,Anthony1970,Okamoto1979}:
\begin{equation}
    \delta_\text{V} = \frac{L_\text{BV}}{L_\text{VV}},
\end{equation}
for the B-V coupling and 
\begin{equation}
    \delta_\text{I} = \frac{L_\text{BI}}{L_\text{II}},
\end{equation}
for the B-SIA coupling.

These factors $\delta_\text{V}$ and $\delta_\text{I}$ give the average number of solute atoms dragged by a vacancy and an SIA, respectively. Note that $L_\text{VV}$, $L_\text{II}$ and $L_\text{BI}$ are systematically positive, while $L_\text{BV}$ may be negative. The off-diagonal coefficient $L_\text{BV}$ determines the sign of the B-V flux coupling. When $\delta_\text{V}<0$, the atom flux is on average opposite to the vacancy flux. When $\delta_\text{V}>0$, vacancies drag solute atoms towards PD sinks through correlated sequences of solute-vacancy exchanges. Note that when the atomic relocation frequency of the FAR mechanism is close to the thermal jump frequency, the flux coupling is reduced in magnitude~\cite{Huang2019}.

\subsubsection{Partial and intrinsic diffusion coefficients}

Even under irradiation, PDs can be considered as dilute species due to their relatively low concentrations. Thus, we choose to ignore the effect of thermodynamic and kinetic interactions between PDs on any diffusion properties. Furthermore, we limit the study to infinitely dilute binary alloys A(B), meaning that we ignore thermodynamic and kinetic interactions between solute atoms B. Therefore, in the A(B) binary alloy with a single type of PD (either V or SIA), the transport coefficients are linear functions of the point defect concentration. To highlight the dependency of the transport coefficients upon the PD concentrations, normalized transport coefficients are introduced, the so-called partial diffusion coefficients~\cite{Marwick1978,Wiedersich1979,Wolfer1983,Nastar2012}, which are independent of PD concentration:
\begin{align}
    &d_\text{AV}=-\frac{L_\text{AA}^{\text{V}} + L_\text{AB}^{\text{V}}}{C_\text{A} C_\text{V}},\,\, d_\text{BV}=-\frac{L_\text{BB}^{\text{V}} + L_\text{BA}^{\text{V}}}{C_\text{B} C_\text{V}}, \nonumber\\
    &d_\text{AI}=\frac{L_\text{AA}^{\text{I}} + L_\text{AB}^{\text{I}}}{C_\text{A} C_\text{I}},\,\, d_\text{BI}=\frac{L_\text{BB}^{\text{I}} + L_\text{BA}^{\text{I}}}{C_\text{B} C_\text{I}}, \nonumber\\
    &d_\text{AV}^c = \frac{L_\text{AA}^{\text{V}}}{C_\text{A} C_\text{V}} - \frac{L_\text{AB}^{\text{V}}}{C_\text{B} C_\text{V}} + d_\text{AV}\frac{1}{\Phi}\xi_\text{VA}, \nonumber\\
    &d_\text{BV}^c = \frac{L_\text{BB}^{\text{V}}}{C_\text{B} C_\text{V}} - \frac{L_\text{BA}^{\text{V}}}{C_\text{A} C_\text{V}} + d_\text{BV}\frac{1}{\Phi}\xi_\text{VB}, \nonumber\\
    &d_\text{AI}^c = \frac{L_\text{AA}^{\text{I}}}{C_\text{A} C_\text{I}} - \frac{L_\text{AB}^{\text{I}}}{C_\text{B} C_\text{I}} + d_\text{AI}\frac{1}{\Phi}\xi_\text{IA}, \nonumber\\
    &d_\text{BI}^c = \frac{L_\text{BB}^{\text{I}}}{C_\text{B} C_\text{I}} - \frac{L_\text{BA}^{\text{I}}}{C_\text{A} C_\text{I}} + d_\text{BI}\frac{1}{\Phi}\xi_\text{IB},
\end{align}
where $\Phi$ is the thermodynamic factor~\cite{Nastar2012}, $\xi_{d\alpha}={(\partial \ln{C_d^\text{eq}})}/{(\partial \ln{C_\alpha})}$, and $C_d^\text{eq}$ is the equilibrium PD concentration. Since multiple-solute and multiple-defect effects are neglected in the dilute-limit approximation, $\Phi$ is equal to 1, and the factors $\xi_{d\alpha}$ are assumed to be zero. Moreover, transport coefficients and intrinsic diffusion coefficients are related by
\begin{equation} \label{eq:D_intrinsic}
    D_{\beta} =  d^c_{\beta \text{V}}C_\text{V} + d^c_{\beta \text{I}}C_\text{I} ,\,\,\,\beta\in\{A,B\}.
\end{equation}
Combining Eqs.\,\eqref{eq:L_total_cluster}, \eqref{eq:C_cluster} and 
\eqref{eq:D_intrinsic}, $D_\beta$ is rewritten as
\begin{equation} \label{eq:D_intrinsic_2}
    D_{\beta} =  d^{c,\text{0}}_{\beta \text{V}}C_\text{V} + d^{c,\text{0}}_{\beta \text{I}}C_\text{I} + L_\text{BB}^\text{mono},\,\,\,\beta\in\{A,B\},
\end{equation}
where $d^{c,\text{0}}_{\beta d}$ is given by 
\begin{align}
    &d^{c,\text{0}}_{\text{A}d} = \left(\frac{L_\text{AA}^{d,\text{pair}}}{C_\text{A} C_d} - \frac{L_\text{AB}^{d,\text{pair}}}{C_\text{B} C_d}\right)C_{\text{B}d}^\text{pair}, \\
    &d^{c,\text{0}}_{\text{B}d} = \left(\frac{L_\text{BB}^{d,\text{pair}}}{C_\text{B} C_d} - \frac{L_\text{BA}^{d,\text{pair}}}{C_\text{A} C_d}\right)C_{\text{B}d}^\text{pair}.
\end{align}
Note that $L_\text{BB}^\text{mono}$ is independent of PD concentrations.

Finally, we express the diffusion coefficients of vacancies ($D_\text{V}$) and SIAs ($D_\text{I}$) in terms of transport coefficients
\begin{equation}
    D_\text{V} = \frac{L_\text{VV}}{C_\text{V}}\,\,\,\text{and}\,\,\,D_\text{I} = \frac{L_\text{II}}{C_\text{I}}.
\end{equation}

\subsection{{Concentration profiles of point defects at sinks}}
The sustained creation of PDs under irradiation and their elimination at sinks leads to a steady-state depleted concentration profile of PDs at sinks. By analogy with the solute RIS, we call it the RIS profile of PDs. We introduce an analytical method to calculate this concentration profile. First, we derive the bulk concentration of PDs at steady state from standard mean-field rate theory~\cite{SIZMANN1978386}. Then, we calculate the steady-state profile by splitting the PD concentration profile in two parts and then integrating the flux of PDs.

\subsubsection{Rate theory}\label{subsubsec:rate_theory}
The concentration of PDs varies under irradiation, mainly due to the production of Frenkel pairs, the mutual recombination between SIA and vacancy, and the elimination of PDs at sinks. We deduce the time-derivative of the bulk concentration of PDs $d=\{{V,I}\}$ under irradiation $C^\text{b}_{d}$ from a classic rate-theory model~\cite{SIZMANN1978386,Russell1984,Schuler2017-2}
\begin{equation} \label{eq:C_V_rate_theory}
\frac{\text{d}C^\text{b}_{d}}{\text{d}t} = \phi - K_R\,C_\text{V}^\text{b}C_\text{I}^\text{b} - K_d (C^\text{b}_{d}-C_{d}^{\text{eq}}),
\end{equation}
where $K_R=({4\pi r_\text{c}}/{\Omega})(D_\text{I}+D_\text{V})$ stands for the SIA-V recombination rate whereas $K_\text{V} = k_\text{V}^2 D_\text{V} $ and $K_\text{I} = k_\text{I}^2 D_\text{I}$ correspond respectively to the elimination rate of vacancies and SIAs at PD sinks.
At PD sinks, we assume that local equilibrium is established and concentrations of PDs correspond to the equilibrium ones. $r_\text{c}$ is the SIA-V recombination radius, usually assumed to be in the same order of magnitude as the lattice parameter $a_0$. $\Omega$ is the atomic volume, $\phi$ the radiation dose rate, i.e., the PD production rate, while $k_\text{V}^2$ and $k_\text{I}^2$ are the sink strength respectively for vacancies and SIAs. We assume that the PD sinks are neutral (i.e., any sink bias is neglected), hence $k_\text{V}^2=k_\text{I}^2=k^2$. Note that $k^2=\sum_s k_s^2$ is the total sink strength, which is summed up over all contributions of the various  PD sinks ($k_s^2$). In steady state, we have
\begin{equation} \label{eq:ideal_sink}
    D_\text{V} (C_\text{V}^\text{b}-C_\text{V}^{\text{eq}}) =  D_\text{I}(C_\text{I}^\text{b}-C_\text{I}^{\text{eq}}).
\end{equation}
Note that, in general, SIAs diffuse much faster than vacancies (i.e., $D_\text{I}\gg D_\text{V}$)~\cite{Was2007}.
Furthermore, the equilibrium SIA concentration $C_\text{I}^{\text{eq}}$ in metals is in general negligible with respect to $C_\text{I}^\text{b}$~\cite{Was2007}. Therefore, the steady-state solution of Eq.\,\eqref{eq:C_V_rate_theory} for $C_\text{V}^\text{b}$ is given by
\begin{equation} \label{eq:V-phi}
    C_\text{V}^\text{b}=\frac{C_\text{V}^\text{eq}}{2}-\dfrac{k^2\Omega}{8\pi r_\text{c}}+\sqrt{\left(\frac{C_\text{V}^\text{eq}}{2}+\dfrac{k^2\Omega}{8\pi r_\text{c}} \right) ^2 + \dfrac{\Omega}{4\pi r_\text{c} }\left(\frac{\phi}{D_\text{V}}\right) }.
\end{equation}
and the general solution for $C_\text{I}^\text{b}$ is obtained from Eq.\eqref{eq:ideal_sink}.
Based on alow temperature expansion  formalism~\cite{Ducastelle1993,Schuler2017-2, Schuler2018, Huang2019}, applied to the infinitely dilute binary alloy A(B) at equilibrium, we have
\begin{equation}\label{eq:CV_eq}
    C_\text{V}^{\text{eq}} = C_\text{V}^{\text{eq,0}} \left[1+ \frac{(Z_{\text{BV}}-Z_{\text{BV}}^0)\overline{C}_\text{B}}{1+(Z_{\text{BV}}-Z_{\text{BV}}^0)C_\text{V}^{\text{eq,0}}}\right],
\end{equation}
where $\overline{C}_\text{B}$ is the nominal concentration of solute atoms B and $C_\text{V}^{\text{eq,0}}$ is the equilibrium concentration in pure metal obtained from the vacancy formation enthalpy $H_\text{V}^\text{f}$ and entropy $S_\text{V}^\text{f}$ by:
\begin{equation}
    C_\text{V}^{\text{eq,0}}=\text{exp}\left( -\frac{H_\text{V}^\text{f}-T\,S_\text{V}^\text{f}}{k_\text{B}\,T} \right).
\end{equation} 
Note that we may neglect $C_\text{V}^\text{eq}$ with respect $C_\text{V}^\text{b}$ in the recombination and sink domains of PDs~\cite{Was2007}. We will take $C_\text{V}^\text{eq}$ into account to define the transition between thermal and sink domains.

In the sink domain, the elimination of PDs at sinks is dominant versus the SIA-V recombination, i.e. $K\gg R$ with $K=K_\text{V}(C_\text{V}^\text{b}-C_\text{V}^{\text{eq}})$ and $R=K_R\,C_\text{V}^\text{b}C_\text{I}^\text{b}$. Therefore, the bulk vacancy concentration $C_\text{V}^\text{b}$ at steady state is proportional to the ratio $\phi/D_\text{V}$ and given by
\begin{equation}\label{eq:CV_sink_dominant}
    C_\text{V}^\text{b} = \frac{1}{k^2}\left(\frac{\phi}{D_\text{V}}\right).
\end{equation} 
In the recombination domain, i.e., when $K\ll R$, the vacancy concentration is proportional to $\sqrt{\phi/D_\text{V}}$, so that we have $C_\text{V}^\text{b}=C_\text{I}^\text{b}$ because Eq.\,\eqref{eq:V-phi} is not valid anymore. In this case, the bulk concentration reads
\begin{equation}\label{eq:CV_recombination_dominant}
    C_\text{V}^\text{b} = \sqrt{\dfrac{\Omega}{4\pi r_\text{c}}\left(\frac{\phi}{D_\text{V}}\right)}.
\end{equation}

\subsubsection{Steady-state RIS profile of point defects} \label{subsubsec:RIS_PD}
\begin{figure}
    \centering
    \includegraphics[width=0.95\linewidth]{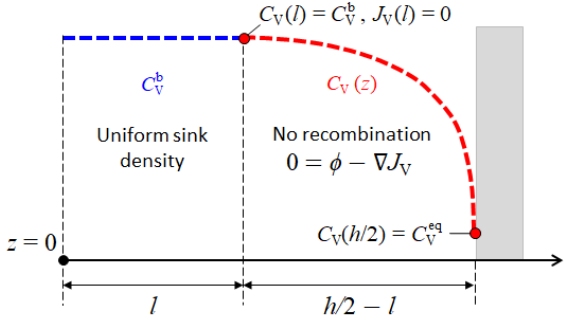}
    \caption{Schema of vacancy concentration profile divided into two regions. In the first region ($0<z<l$), the vacancy concentration is assumed to be uniform in space and given by a mean-field kinetic approach. In the second region ($l<z<h/2$), the PD recombination reactions are neglected and the vacancy concentration is given by $\div \bm{J}_\text{V}=\phi$.}
    \label{fig:CV_schema}
\end{figure}

Homogeneous mean-field rate theory does not provide the concentration profiles of PDs at sinks that form  under sustained irradiation. From the continuity equation,
we express the elimination rate of PDs as a divergence of the vacancy flux $\bm{J}_\text{V}$ and the SIA flux $\bm{J}_\text{I}$:
\begin{align}
     \label{eq:CV_diffusion_flux}
     \frac{\partial C_\text{V}}{\partial t}\, &=\phi - K_R C_\text{V} C_\text{I} - \div{\bm{J}_\text{V}} , \\
     \label{eq:CI_diffusion_flux}
     \frac{\partial C_\text{I}}{\partial t}\, &=\phi - K_R C_\text{V} C_\text{I} - \div{\bm{J}_\text{I}}.
\end{align}
To compute $\bm{J}_\text{V}$ and $\bm{J}_\text{I}$, we assume $\grad{\mu_\text{V}}$ and $\grad{\mu_\text{I}}$ to be the dominant driving forces compared to $\grad{\mu_\text{A}}$ and $\grad{\mu_\text{B}}$. Moreover, we assume that the chemical potential of PDs does not depend on the solute concentration. Thus, we have
\begin{align}
    &\bm{J}_\text{V} = -\frac{L_\text{VV}}{k_\text{B} T}\grad{\mu_\text{V}} = -D_\text{V}\grad{C_\text{V}}, \\
    &\bm{J}_\text{I} = -\frac{L_\text{II}}{k_\text{B} T}\grad{\mu_\text{I}} = -D_\text{I}\grad{C_\text{I}}.
\end{align}
We present the calculation of the PD concentration profile in a one-dimensional symmetric system delimited by two  planar sinks parallel to each other, as illustratd in Fig. \ref{fig:CV_schema}. These planar sinks may represent ideal surfaces, grain-boundaries, or interfaces. We define the PD concentration profile along the coordinate axis $(Oz)$ perpendicular to the planar sinks. {We assume that $D_\text{V}$ and $D_\text{I}$ do not depend on the spatial coordinates, i.e. they do not vary along the solute RIS profile.} At steady state and along the axis $(Oz)$, the diffusion equations of SIAs and vacancies lead to the partial differential equations: 
\begin{align}
     \label{eq:CV_diffusion}
     0\, &=\phi - K_R C_\text{V} C_\text{I} + D_\text{V}\frac{\partial^2 C_\text{V}}{\partial z^2}, \\
     \label{eq:CI_diffusion}
     0\, &=\phi - K_R C_\text{V} C_\text{I} + D_\text{I}\frac{\partial^2 C_\text{I}}{\partial z^2},
\end{align}
where $C_\text{V}(z)$ and $C_\text{I}(z)$ are respectively the local concentrations of vacancies and SIAs at coordinate $z$.
At any position $z$, these local concentrations are related to each other by the steady-state relation $D_\text{V} \left[C_\text{V}(z)-C_\text{V}^\text{eq}\right]=D_\text{I} C_\text{I}(z)$, as demonstrated in~\ref{subsec:C_V and C_I}. {Since $C_\text{V}$ and $C_\text{I}$ are related, in the following we consider only the spatial variation of $C_\text{V}$.} Assuming that $D_\text{I}\gg D_\text{V}$, Eq.\,\eqref{eq:CV_diffusion} is rewritten as follows: 
\begin{equation}\label{eq:CV_diffusion_2}
    \frac{\partial^2 C_\text{V}(z)}{\partial z^2}=-\frac{\phi}{D_\text{V}}+ \frac{4\pi r_\text{c}}{\Omega}C_\text{V}(z)\left[ C_\text{V}(z)-C_\text{V}^\text{eq}\right].
\end{equation}
Note that, for the time being, a general analytical solution of  Eq.\,\eqref{eq:CV_diffusion_2} does not exist~\cite{Davis1962,Lam1974,Rauht1981} 
Close to a planar sink and if we neglect the mutual recombination reactions between PDs (i.e. $r_\text{c}=0$), there is a simple analytical solution of the PD concentration profile~\cite{Martinez2018}).  As explained in Ref.\,\cite{Martinez2018}, the solution of Eq.\,\eqref{eq:CV_diffusion_2} with $r_\text{c}=0$, reads:
\begin{equation}\label{eq:CV_prl}
    C_\text{V}(z)=-a(z^2-\frac{h^2}{4}) + C_\text{V}^\text{eq},
\end{equation}
where $a=\phi/(2D_\text{V})$, $h$ is the average spacing between planar sinks. and the position of the origin of axis ($0z$) is chosen to be at the mid-point between two planar sinks (see Fig.\,\ref{fig:CV_schema}). 

Close to a PD sink, ignoring the recombination reactions should be a reasonable hypothesis, because locally concentrations of PDs are very low. Hence their probability of recombination, that is  proportional to the square of the PD concentrations, should be very low as well.
Therefore, we split the PD concentration profiles in two regions: a bulk region far from sinks in which concentrations are uniform, and a sink region in which we account for the $z$-variation of the PD concentration profile (cf. Fig.\,\ref{fig:CV_schema}). The $z$ coordinate of the bulk region ranges between $0$ and $l$, whereas the $z$-coordinate of the sink region ranges from $l$ to $h/2$, $h$ being the distance between the planar sinks.  In the bulk region, $C_\text{V}(z)$ is constant and equal to the steady-state bulk concentration $C_\text{V}^\text{b}$ (see Eq.\,\eqref{eq:C_V_rate_theory}). From Eq.\,\eqref{eq:CV_prl}, we deduce the vacancy concentration profile $C_\text{V}(z)$, with $l\leq z\leq h/2$. In order to ensure the continuity of the vacancy concentration and its spatial derivative (i.e., the vacancy flux), we apply the boundary conditions
\begin{equation}
    C_\text{V}(l^-) = C_\text{V}(l^+),\,\,\frac{\partial C_\text{V}}{\partial z}(l^-) =\frac{\partial C_\text{V}}{\partial z}(l^+)=0. 
\end{equation}
At PD sinks, the vacancy concentration corresponds to the equilibrium one:
\begin{equation}
    C_\text{V}(\frac{h}{2}) =C_\text{V}^\text{eq}.
\end{equation}
The solution is then given by
\begin{equation}\label{eq:CV_two_regimes}
    C_\text{V}(z)= 
    \begin{cases}
        C_\text{V}^\text{b}, &\text{$0\leq z<l$}; \\
        -a\left( z-l\right)^2 + C_\text{V}^\text{b}, & \text{$l\leq z\leq h/2$},
    \end{cases}
\end{equation}
where the characteristic distance $l$ is defined as
\begin{equation}
    l = \frac{h}{2}-\sqrt{\frac{C_\text{V}^\text{exc}}{a}},
\end{equation}
where $C_\text{V}^\text{exc} = C_\text{V}^\text{b} - C_\text{V}^\text{eq}$ corresponds to the vacancy excess concentration with respect to the equilibrium one. Note that the characteristic distance $l$ depends on the interplanar distance $h$ and $C_\text{V}^\text{exc}$. Both quantities are related to the microstructure. {The interplanar distance $h$ determines the sink strength of the parallel planar sinks~\cite{Nichols1978}: 
\begin{equation}\label{eq:interplanr_sink_strength}
    k^2=\frac{8}{h^2}.
\end{equation}
In case there is no other PDs sinks in the system, this sink strength fully determines the bulk concentration of vacancy, $C_\text{V}^\text{b}$. Note that in case there are other sinks, in addition to the local sink strength of the planar sinks, $C_\text{V}^\text{b}$ depends on the overall sink strength of the other PDs sinks of the microstructure.}
If $l<0$, Eq. \ref{eq:CV_two_regimes} is no longer appropriate, because the PDs planar sinks are so close that it is not possible to introduce a bulk region with uniform concentrations. In this case, we set $l=0$, and the obtained PD concentration profile is given by Eq. \ref{eq:CV_prl}.

Using the Gibbs formalism of interface excess quantities, we define the vacancy concentration excess at sinks by the following integral:
\begin{equation}  \label{eq:segregation_amount}
    S_\text{V} =  \int_0^{h/2}\left[ C_\text{V}(z) - C_\text{V}(0) \right]\text{d}z.
\end{equation}
We obtain from Eq.\,\eqref{eq:CV_two_regimes} and Eq.\,\eqref{eq:segregation_amount} that:
\begin{equation} \label{eq:SV}
    S_\text{V} = -\frac{\left(C_\text{V}^\text{b}-C_\text{V}^\text{eq}\right)^\frac{3}{2}}{3\sqrt{a}}.
\end{equation}
As expected, $S_\text{V}$ is always negative. 
Note that the latter depends on the PD recombination reactions through the variation of $C_\text{V}^\text{b}$ with $R$. 
Therefore, as stated in Sec.\,\ref{subsubsec:rate_theory}, we cannot ignore the recombination reactions, 
unless the recombination rate ($R$) is negligible with respect to the PD elimination rate at sinks ($K$).

In the sink domain, i.e., $K\gg R$, we have
\begin{equation} \label{eq:SV_elimination_dominant}
    S_\text{V} = -\frac{\sqrt{2}}{3\left( k^2 \right)^{3/2}}\left(\frac{\phi}{D_\text{V}}\right).
\end{equation}
Therefore, $S_\text{V}$ is proportional to the ratio $\phi/D_\text{V}$, and it decreases with the sink strength $k^2$.

In the recombination domain, i.e. $R\gg K$, we have
\begin{equation} \label{eq:SV_recombination_dominant}
    S_\text{V} = -\frac{1}{6} \left(\frac{\Omega}{\pi r_\text{c}}\right)^{3/4}\left( \frac{\phi}{D_\text{V}} \right)^{1/4}.
\end{equation}
Thus, $S_\text{V}$ is proportional to $(\phi/D_\text{V})^\frac{1}{4}$, and it is independent of $k^2$.

\subsection{Radiation induced segregation of solute atoms}\label{subsec:RIS_solute}
From the vacancy RIS profile and the RIS factor expressed as functions of the local PD and solute concentrations, we derive an analytical expression of the solute RIS profile.

\subsubsection{Local concentration-dependent RIS factor {$\alpha$}}\label{subsubsec:alpha}
According to Wiedersich~\cite{Wiedersich1979,Nastar2012,Ardell2016}, at steady-state, the concentration gradients of solutes and vacancies near an ideal sink are related by
\begin{align} \label{eq:grad_CB_CV}
    \grad C_\text{B}=-\alpha(z){\grad C_\text{V}},
\end{align} 
with the RIS factor 
\begin{equation} \label{eq:alpha}
    \alpha(z) = \frac{d_\text{AI}d_\text{AV}C_\text{A} C_\text{B}}{d_\text{AI}D_\text{B}C_\text{A} + d_\text{BI}D_\text{A}C_\text{B}}\alpha_s,
\end{equation}
and 
\begin{equation}
    \alpha_s = \frac{d_\text{BI}}{d_\text{AI}}-\frac{d_\text{BV}}{d_\text{AV}}.
\end{equation}
$\alpha_s$ determines the sign of the RIS factor $\alpha$, which in turn determines the sign of RIS. In a dilute binary alloys A(B), the partial diffusion coefficients $d_\text{BI}$, $d_\text{AI}$ are systematically positive, while $d_\text{AV}$ is systematically negative. On the other hand, $d_\text{BV}$ is either positive or negative. The off-diagonal coefficient $L_\text{BV}$ determines the sign of $d_\text{BV}$. Concerning the sign of $\alpha_s$, we consider two cases. When $d_\text{BV}$ is positive, i.e., vacancies drag solute atoms, $\alpha_s$ is positive and RIS leads to solute enrichment around sinks. When $d_\text{BV}$ is negative, the sign of $\alpha_s$ depends on the relative amplitude of the partial diffusion coefficient ratios $d_\text{BI}/d_\text{AI}$ and $d_\text{BV}/d_\text{AV}$. Note that a steady-state gradient of solute concentration can only be established if a backward diffusion opposes the solute gradient resulting from flux coupling. The rate of this backward reaction is governed by the intrinsic diffusion coefficients $D_\text{A}$ and $D_\text{B}$ that appear in the denominator of $\alpha (z)$.

The local RIS factor $\alpha$ depends on the $z$ coordinate through the variation of the local concentration $C_\text{B}$ and $C_\text{V}$ with $z$ (see Eq.\,\eqref{eq:alpha}). In order to analyze the variation of $\alpha$ with the local concentrations of vacancies and solute atoms, we rewrite $\alpha$ by making explicit its variation with $C_\text{B}$ and $C_\text{V}$. {Note that we neglect PD concentration with respect to the solute and solvent concentration, i.e., $C_\text{A}\equiv 1-C_\text{B}$.} We deduce from  Eqs.\,\eqref{eq:D_intrinsic_2},\,\eqref{eq:ideal_sink} and \eqref{eq:alpha} that
\begin{equation} \label{eq:alpha_1_2}
    \alpha(z) = \frac{\alpha_1 C_\text{B}(z)}{C_\text{V}(z) + \alpha_2},
\end{equation}
where
\begin{align}
    \label{eq:alpha_1}&\alpha_1 = \frac{\alpha_0}{\alpha_\text{V} + \alpha_\text{I} },\\
    \label{eq:alpha_2}&\alpha_2 = \frac{\alpha_{\text{mono}} - \alpha_\text{I} C_\text{V}^\text{eq} }{\alpha_\text{V} + \alpha_\text{I}},
\end{align}
with
\begin{align}
    &\alpha_0 = \alpha_s\,C_\text{A} d_\text{AI}d_\text{AV} ,\nonumber \\
    &\alpha_\text{V} = C_\text{A} d_\text{AI}d_\text{BV}^{c} + C_\text{B} d_\text{BI} d_\text{AV}^{c},\nonumber \\
    &\alpha_\text{I} = (C_\text{A} d_\text{AI}d_\text{BI}^{c} + C_\text{B} d_\text{BI} d_\text{AI}^{c})D_\text{V}/D_\text{I},\nonumber \\
    &\alpha_\text{mono} =(C_\text{A} d_\text{AI} + C_\text{B} d_\text{BI}) L_\text{BB}^{\text{mono}} \nonumber.
\end{align}
Therefore, $\alpha$ decreases with $C_\text{V}$. The RIS factors $\alpha_1$ and $\alpha_2$ are independent of the PD concentrations. Instead, they vary with $C_\text{B}$ because the partial diffusion coefficients $d_{ij}$ depend on $C_\text{B}$. However, along a RIS profile, the relative variation of the solute concentration is a lot smaller than that of the PD concentration. Hence, we assume that $\alpha_1$ and $\alpha_2$ do not vary along the RIS profile, and we compute these coefficients at $C_\text{B}$ equal to the nominal solute concentration. {This assumption is later justified by a comparison between the solute RIS profile obtained from our analytical approximation and the exact solution (cf. Fig.\,\ref{fig:CB_Fe-X_profiles}).}

\subsubsection{Steady-state RIS profile of solute atoms}
Following Eq.\,\eqref{eq:CV_two_regimes}, we deduce the vacancy concentration gradient $\grad C_\text{V}$. Then, from Eqs.\,\eqref{eq:grad_CB_CV} and \eqref{eq:alpha_1_2}, we obtain the solute concentration gradient 
\begin{equation} \label{eq:grad_CV}
    \frac{\grad C_\text{B}}{C_\text{B}}(z)=
    \begin{cases}
        0, & \text{$0\leq z<l$}; \\
        -\frac{2\,\alpha_1(z-l)}{\left( z-l\right)^2 - b^2}, & \text{$l\leq z\leq h/2$},
    \end{cases}
\end{equation}
with $b^2=(C_\text{V}^\text{b} + \alpha_2)/a$.

We determine $\alpha_1$ and $\alpha_2$ from the nominal solute concentration, $\overline{C}_\text{B}$. We derive the concentration profile of the solute atoms by integrating Eq.\,\eqref{eq:grad_CV} with respect to $z$.
We deduce the constants of integration from
the following boundary conditions
\begin{align}
    \label{eq:CB_continuity}
    & C_\text{B}(l^-) = C_\text{B}(l^+), \\
    \label{eq:mass_conservation_condition}
    & \int_0^{h/2}C_\text{B}(z)dz  = \frac{h}{2}\,\overline{C}_\text{B}.
\end{align}
These conditions ensure the continuity of the solute concentration profile $C_\text{B}$ at $z=l$ and the mass conservation of the solute atoms along the RIS profile. 

{Note that in this study we neglect the equilibrium segregation of solutes resulting from the interaction of solutes with the sink~\cite{Foiles1989,Menyhard1994,Creuze2000}. This thermodynamic property may strongly modify the solute concentration over the first two or three atomic planes of the sink~\cite{Foiles1989,Menyhard1994}. Its amplitude and width (generally less than $1$ nm) vary with the temperature, the chemical nature of solute atoms, and the nature of the sink. A quantitative investigation of this phenomenon would require a detailed knowledge of the structure of the sink as well as the solute segregation energies at different atomic sites near the sink. Nevertheless, the total amount of solute RIS segregation, as well as the average width of the RIS profiles (spreading over a few tens of nanometers~\cite{Was2007}) should not be much affected by the equilibrium segregation.}

According to the boundary conditions, i.e., Eqs\,\eqref{eq:CB_continuity} and \eqref{eq:mass_conservation_condition}, we obtain
\begin{equation}\label{eq:CB_two_regimes}
    C_\text{B}(z)=
    \begin{cases}
        K_1\,b^{-2\alpha_1}, & \text{$0\leq z<l$}; \\
        K_1\left[ b^2-(z-l)^2\right]^{-\alpha_1}, & \text{$l\leq z\leq h/2$},
    \end{cases}
\end{equation}
with 
\begin{equation} \label{eq:K1}
    K_1 = \frac{h}{2}\frac{\overline{C}_\text{B}}{lb^{-2\alpha_1}+\int_0^{(h/2)-l}\left( b^2-z^2\right)^{-\alpha_1}\text{d}z}.
\end{equation}
Note that there is no simple analytical expression of the integral $I=\int_0^{(h/2)-l}\left( b^2-z^2\right)^{-\alpha_1}\text{d}z$. Nevertheless, we can calculate it from the hypergeometric function $_2F_1$~\cite{Andrews1999} (presented in~\ref{subsec:hypergeometric_function} ).

Similarly to Eq.\,\eqref{eq:SV}, we define the total amount of solute atoms segregated at sinks as
the solute concentration excess 
$S_\text{B}=(h/2)\left(\overline{C}_\text{B}-C_\text{B}(0)\right)$. It writes
\begin{equation} \label{eq:SB_two_regimes_2}
    S_\text{B} = \frac{h}{2}\left(\overline{C}_\text{B} - K_1\,b^{-2\alpha_1}\right). 
\end{equation}

In cases where we cannot ignore the FAR mechanism, the diffusion of isolated solute atoms ($L_\text{BB}^\text{mono}$) and thus, $\alpha_2$ are not negligible. $\alpha_2$ increases with the FAR frequency, which in turn decreases the RIS of solute atoms. In the extreme case where
$\alpha_2\gg C_\text{V}^\text{b}$, we obtain for $0<z<(h/2)-l$:
\begin{equation}
    z^2 < \left(\frac{h}{2}-l\right)^2 < C_\text{V}^\text{b} / a \ll b^2.
\end{equation}
In this case, $I \simeq [(h/2)-l]b^{-2\alpha_1}$, $K_1 = \overline{C}_\text{B}\,b^{2\alpha_1}$, and the amount of segregated solute is zero ($S_\text{B}=0$).   

On the contrary, if we ignore FAR, $\alpha_2\ll C_\text{V}^\text{b}$. If we also neglect $C_\text{V}^\text{eq}$,  $b = (h/2)-l$ and we obtain
\begin{align} \label{eq:I}
    I & = \int_0^{b}(b^2-z^2)^{-\alpha_1}\text{d}z, \nonumber \\
      & = b^{-2\alpha_1 + 1}I_{\alpha_1},
\end{align}
with 
\begin{equation} \label{eq:I_alpha_1}
    I_{\alpha_1} = \int_0^{1}(1-z^2)^{-\alpha_1}\text{d}z.
\end{equation} 
Note that $I_{\alpha_1}$ is positive and only depends on the RIS factor $\alpha_1$. Moreover, it is larger than 1 if the solute RIS is positive (i.e., $\alpha_1>0$), and smaller than 1 in the opposite case. Then, after Eqs.\,\eqref{eq:K1}, \eqref{eq:SB_two_regimes_2}, and \eqref{eq:I}, we may approximate $S_\text{B}$ as follows
\begin{align}
   \label{eq:SB_CV_irr_dominant_0}S_\text{B} & = \frac{h}{2}\,\overline{C}_\text{B} \frac{I_{\alpha_1}-1}{I_{\alpha_1}+\frac{l}{({h}/{2})-l}}, \\
       \label{eq:SB_CV_irr_dominant}& = \frac{h}{2}\,\overline{C}_\text{B} \frac{I_{\alpha_1}-1}{I_{\alpha_1}-1+\frac{h}{2}\sqrt{\frac{a}{C_\text{V}^\text{b}}}}.
\end{align}
According to Eq.\,\eqref{eq:SB_CV_irr_dominant_0}, $S_\text{B}$ is positive if the solute RIS factor is positive, and negative otherwise. 

Besides, following Eqs.\eqref{eq:SV} and \eqref{eq:SB_CV_irr_dominant}, we obtain a direct relationship between $S_\text{B}$ and $S_\text{V}$
\begin{equation} \label{eq:SB_SV}
   S_\text{B} = \frac{h}{2}\,\overline{C}_\text{B} \frac{I_{\alpha_1}-1}{I_{\alpha_1}-1-\frac{h}{2}{\frac{C_\text{V}^\text{b}}{S_\text{V}}}}.
\end{equation}
We observe that the amount of solute RIS, $S_\text{B}$, is directly related to $\alpha_1$ and $S_\text{V}/C_\text{V}^\text{b}$.
In the denominator, $C_\text{V}^\text{b}$ is the signature of the backward diffusion opposing the RIS solute concentration gradient. This backward diffusion is the reason why solute RIS ($S_\text{B}$), unlike PD RIS ($S_\text{V}$), is not systematically governed by the ratio $\phi/D_\text{V}$.

In the sink domain ($K\gg R$), we obtain from Eq.\,\eqref{eq:CV_sink_dominant} and Eq.\,\eqref{eq:SB_CV_irr_dominant} that
\begin{equation} \label{eq:SB_elimination_dominant}
    S_\text{B} = \frac{h}{2}\,\overline{C}_\text{B} \frac{I_{\alpha_1}-1}{I_{\alpha_1}-1+\frac{h}{2}\sqrt{\frac{k^2}{2}}}.
\end{equation}
In this case, $S_\text{B}$ is independent of the ratio $\phi/D_\text{V}$, whereas it decreases with $k^2$, as already shown in Ref.\,\cite{Martinez2018}. Thus, at fixed $k^2$, if we neglect the small variation of $\alpha_1$  with the dose rate $\phi$, the solute RIS amount is independent of $\phi$~\cite{Martinez2018}. Besides, $S_\text{B}$ varies with temperature through the variation of $\alpha_1$ with temperature. 
Note that the present expression of $S_\text{B}$ is not exactly the same as the one published in Ref.~\cite{Martinez2018}, because here we do not assume that the RIS factor is independent of solute and PD concentrations.  

In the recombination domain  ($K\ll R$), we obtain from Eq.\,\eqref{eq:CV_recombination_dominant} and Eq.\,\eqref{eq:SB_CV_irr_dominant} that
\begin{equation} \label{eq:SB_recombination_dominant}
    S_\text{B} = \frac{h}{2}\,\overline{C}_\text{B} \frac{I_{\alpha_1}-1}{I_{\alpha_1}-1+\frac{h}{2}\left(\frac{\Omega}{\pi r_\text{c}  }\right)^{-1/4}\left(\frac{\phi}{D_\text{V}}\right)^{1/4}}
\end{equation}
$S_\text{B}$ is then governed by the ratio $\phi/D_\text{V}$ as well as by the RIS factor $\alpha_1$. Moreover, it decreases with the dose rate $\phi$.

\subsection{Interplay between RIS and the microstructure}
According to the analytical derivation, the RIS segregation profiles between parallel planar sinks depend upon the spacing $h$ between the parallel planar sinks, and the total sink strength $k^2=\sum_s k^2_s$ of the overall microstructure including the local parallel planar sink strength (Eq.\,\eqref{eq:interplanr_sink_strength}). Such a modeling of the PD sink population allows for the investigation of the RIS profile of a local sink interacting with the overall microstructure. 
Note that our analytical model of RIS could be easily extended to other local sinks, such as dislocation lines $k^2_\text{line}$ and dislocation loops $k^2_\text{loop}$. 

In the following, for the sake of simplicity, we consider parallel sink planes are the major PD sinks. This means that we ignore the contributions of other types of sinks, and relate the inter-planar distance to the total sink strength: $h=\sqrt{8/k^2}$.

\section{Results} \label{Sec:Results_and_discussions}

We apply the above RIS models to the specific case of Fe-based dilute alloys. We start this section with a brief presentation
of these alloys from a perspective of their PD energy properties. Then we present a parametric study of the variation of steady-state vacancy concentration with temperature, radiation flux, and sink strength, with the aid of 2-D maps.
We extend this parametric approach to the solute diffusion coefficients, the flux coupling and RIS factors, and the vacancy and solute RIS profiles. 
Note that in the temperature--radiation flux maps, the sink strength is set to $k^2=5\times 10^{14}\,\text{m}^{-2}$ and the recombination radius $r_\text{c}$ is set to $\sqrt{3}\,a_0$ unless otherwise specified.

\subsection{DFT energy database of dilute Fe-based alloys}
\begin{table}
\caption{\label{tab:solute-PD_binding_Fe-X}%
\textit{Ab initio} solute-PD binding energies (in eV) of Fe alloys obtained in Ref.\,\cite{Messina2014} for mixed B-I dumbbell configuration, and Ref.\,\cite{Messina2019} for 1-NN and 2-NN B-V pairs configurations. Negative energies stand for attractive interactions. 
}
\centering
    \begin{ruledtabular}
    {\begin{tabular}{c c c c c c c}
    Configuration & Fe-P & Fe-Mn & Fe-Cr & Fe-Si & Fe-Ni & Fe-Cu \\
    \hline
    Mixed B-I    & -1.03 & -0.56 & -0.05 & +0.00 & +0.19 & +0.38 \\
    1-NN B-V     & -0.38 & -0.17 & -0.06 & -0.30 & -0.10 & -0.26 \\
    2-NN B-V     & -0.27 & -0.11 & -0.01 & -0.11 & -0.21 & -0.17 \\
    \end{tabular}
    }
    \end{ruledtabular}
\end{table}

The vacancy formation enthalpy $H_\text{V}^\text{f}$ and entropy $S_\text{V}^\text{f}$ in pure iron are respectively set to 2.18\,eV and 4.1 $k_\text{B}$, and the lattice parameter $a_0$ to 2.831\,\AA~ {according to previous DFT calculations}~\cite{Messina2014}. The \textit{ab initio} solute-PD binding energies, migration energies, and jump frequency prefactors are found in~Ref.\,\cite{Messina2014} for the vacancy diffusion mechanism, and in Ref.\,\cite{Messina2019} for the dumbbell diffusion mechanism. 

{The computation of the RIS factors in these alloys has shown that the general flux coupling behavior is largely governed by the short-range {thermodynamic} interaction between PDs and solute atoms~\cite{Messina2014}. We list in Tab.\,\ref{tab:solute-PD_binding_Fe-X} the binding energy values of the mixed dumbbell and the 1- and 2-NN solute-vacancy pairs. In the six binary alloys, the 1- and 2-NN solute-vacancy binding energies are negative, i.e., the solute atoms are attracted by the vacancy. Moreover, Cr has a very weak interaction with vacancies compared with the other solute atoms. {Concerning the SIAs, the most stable configuration is the dumbbell one. {Based on} the binding energies of the solute-Fe mixed dumbbells,} we can divide the Fe-based dilute alloys into two groups: {those with} stable (P, Mn, Cr) and non-stable (Si, Ni, Cu) mixed dumbbells. 

Note that the kinetic behaviors of the solute atoms forming non-stable mixed dumbbells (Si, Ni, Cu) should be predominantly controlled by the vacancy mechanism~\cite{Messina2019}. In addition, the values of the 1-NN an 2-NN solute-vacancy binding energies in Fe-Si, Fe-Ni, and Fe-Cu alloys are close. Therefore, the kinetic properties of the solute atoms are expected to be similar in these alloys. On the other hand, for the group of solutes forming stable mixed dumbbells (P, Mn, Cr), the values of solute-PD binding energies {cover a wider range}, suggesting that the kinetic properties can be very different in Fe-P, Fe-Mn, and Fe-Cr alloys. 
}

\subsection{Bulk vacancy concentration at steady state}
\begin{figure}
    \centering
    \includegraphics[width=1.0\linewidth]{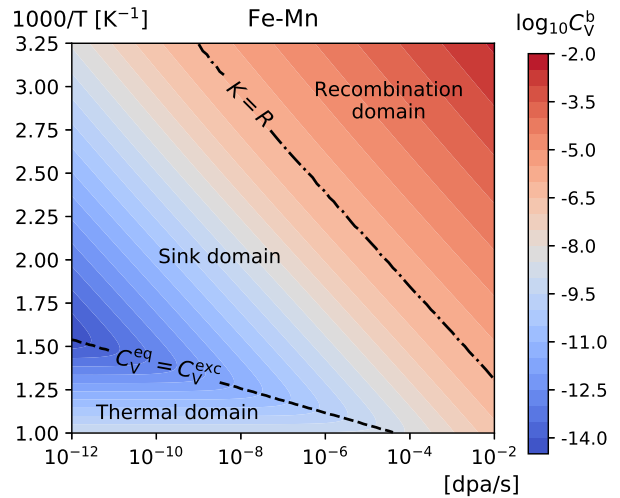}
    \caption{Bulk vacancy concentration $C_\text{V}^\text{b}$ as a function of dose rate (in dpa/s) and inverse temperature (in K$^{-1}$) for the dilute Fe-Mn alloy. The nominal solute concentration $\overline{C}_\text{Mn}$ is set to 1\,at.\% and the sink strength $k^2$ is set to $5\times 10^{14}\,\text{m}^{-2}$. {The main trends of $C_\text{V}^\text{b}$ with $T$ and $\phi$ are similar for the other investigated Fe alloys.}}
    \label{fig:CV_Fe-X}
\end{figure}
As shown in Sec.\,\ref{subsec:RIS_solute}, the RIS of solute atoms is related to the bulk vacancy concentration. Thus, the kinetic domains of solute RIS are similar to the PD kinetic domains. Therefore, we compute the steady-state bulk  concentration of vacancies with respect to temperature, radiation flux, and sink strength.

First, we compute the variation of the bulk vacancy concentration, $C_\text{V}^\text{b}$, with temperature ($T$) and dose rate ($\phi$). In Fig. \ref{fig:CV_Fe-X}, the result of Fe-Mn system is represented in the form of a $\phi$--$T$ map, in which the colors indicate the amplitude of $C_\text{V}^\text{b}$. This map can be divided into three domains: thermal domain when the bulk vacancy concentration is lower than twice the equilibrium vacancy concentration, i.e., the effect of irradiation is negligible; sink domain for $K>R$; recombination domain for $R>K$. 

As expected from the analytical results  (cf. Eqs.\,\eqref{eq:CV_sink_dominant} and \eqref{eq:CV_recombination_dominant}), $C_\text{V}^\text{b}$ decreases with $T$, whereas it increases with $\phi$. In the recombination domain, $C_\text{V}^\text{b}$ increases linearly with $\phi$. In both the recombination and sink domains, each level line corresponds to a fixed value of $\phi/D_\text{V}$.
In the thermal domain, the level lines are horizontal because $C_\text{V}^\text{b}$ is  close to the equilibrium vacancy concentration $C_\text{V}^\text{eq}$, and independent of $\phi$. 

{We find that the main trends of $C_\text{V}^\text{b}$ with $T$ and $\phi$ are similar for the other investigated Fe alloys (not presented), with only slight variations of the extent of the kinetic domains. Therefore, we conclude that the solute effect on the bulk concentration of vacancies is negligible.}

\subsection{Solute diffusion}
\begin{figure*}
    \centering
    \includegraphics[width=1.0\linewidth]{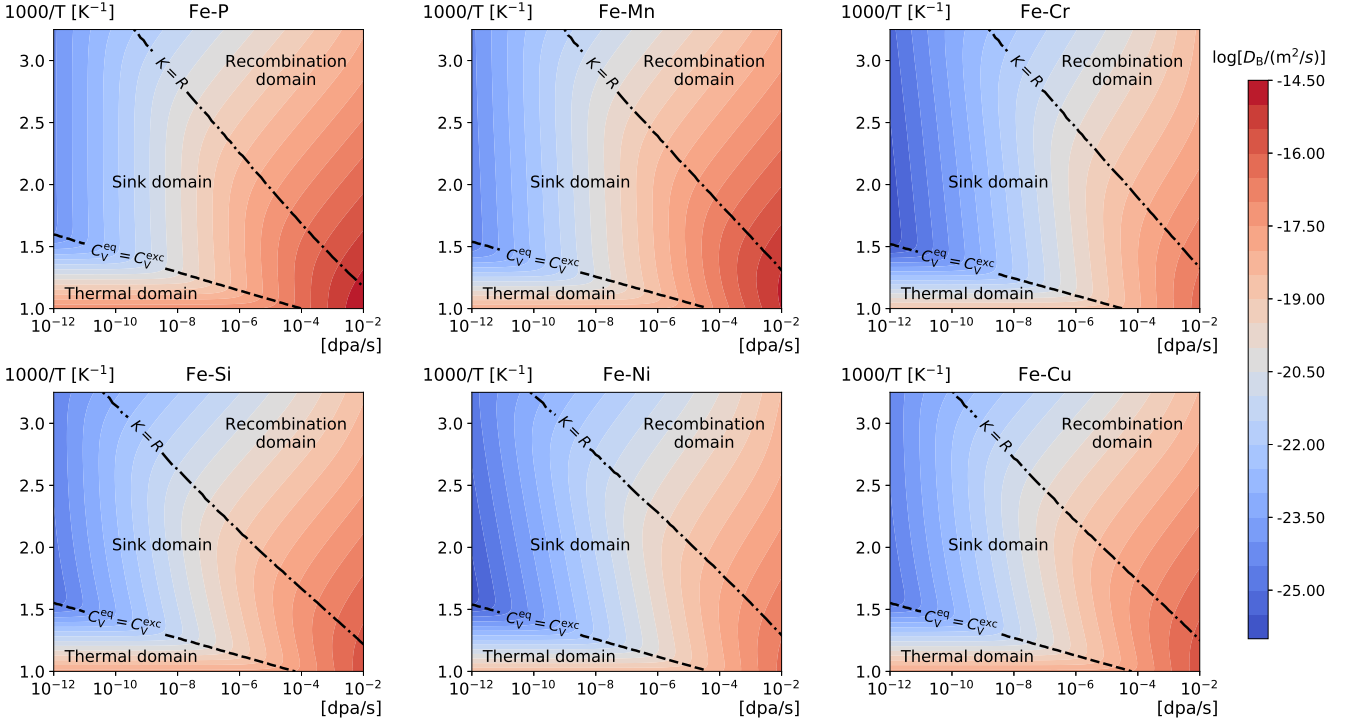}
    \caption{Solute diffusion coefficient $D_\text{B}$ as a function of dose rate (in dpa/s) and inverse temperature (in K$^{-1}$) for  several dilute binary Fe-based alloys. The nominal solute concentration $\overline{C}_\text{B}$ is set to 1\,at.\% and the sink strength $k^2$ is set to $5\times 10^{14}\,\text{m}^{-2}$.}
    \label{fig:DB_Fe-X}
\end{figure*}
We compute the intrinsic solute diffusion coefficient, $D_\text{B}$, which is equivalent in a dilute alloy to the solute tracer diffusion coefficient. The $\phi$--$T$ maps of $D_\text{B}$ are presented in Fig.\,\ref{fig:DB_Fe-X}. For the six alloys, we recover the  three kinetic domains of $C_\text{V}^\text{b}$.
{Over most irradiation conditions of interest, $C_\text{B}\gg C_\text{V}^\text{b}$ (see Fig.\,\ref{fig:CV_Fe-X}). In this case, the solute diffusion coefficient varies linearly with the bulk vacancy concentration (cf. Eq.\,\eqref{eq:D_intrinsic}), provided the effect of FAR is negligible. Therefore, we expect the same kinetic domains, except in the low-temperature and high-flux domain where FAR may affect the solute diffusion properties and $C_\text{B}$ has the same order of magnitude than $C_\text{V}^\text{b}$.}
  
$D_\text{B}$ increases with temperature in the recombination and thermal domains, though the increasing rate is different in the two domains. In the sink domain, $D_\text{B}$ is nearly $T$-independent.
As for the effect of the radiation flux, $D_\text{B}$ increases with $\phi$ except in thermal domain. 
Similarly to $C_\text{V}^\text{b}$, the solute effect on the main trends is weak.  

\subsection{Flux coupling}
\begin{figure}
    \centering
    \includegraphics[width=1.0\linewidth]{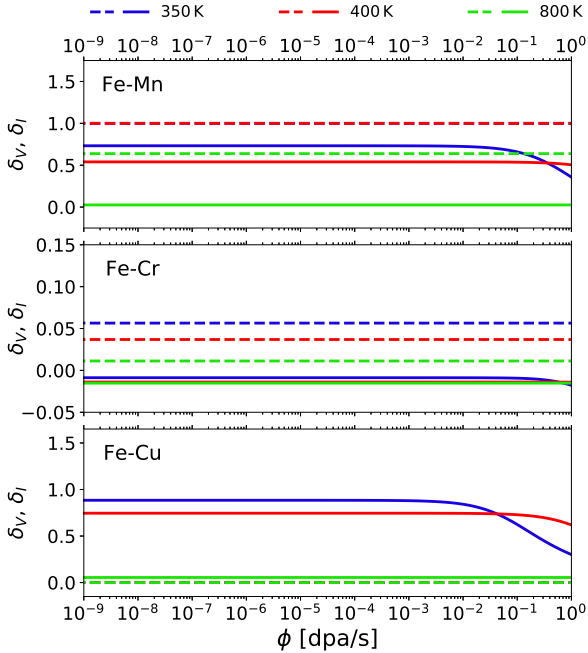}
    \caption{Flux coupling coefficients for Fe-Mn, Fe-Cr and Fe-Cu alloys, mediated by vacancies (solid lines) and SIAs (dashed line), as functions of dose rate. The results are obtained for three different temperatures: 350\,K, 450\,K and 800\,K. The nominal solute concentration $\overline{C}_\text{B}$ is set to 1\,at.\% and the sink strength $k^2$ is set to $5\times 10^{14}\,\text{m}^{-2}$.}
    \label{fig:delta_Fe-X}
\end{figure}

\begin{figure*}
    \centering
    \includegraphics[width=1.0\linewidth]{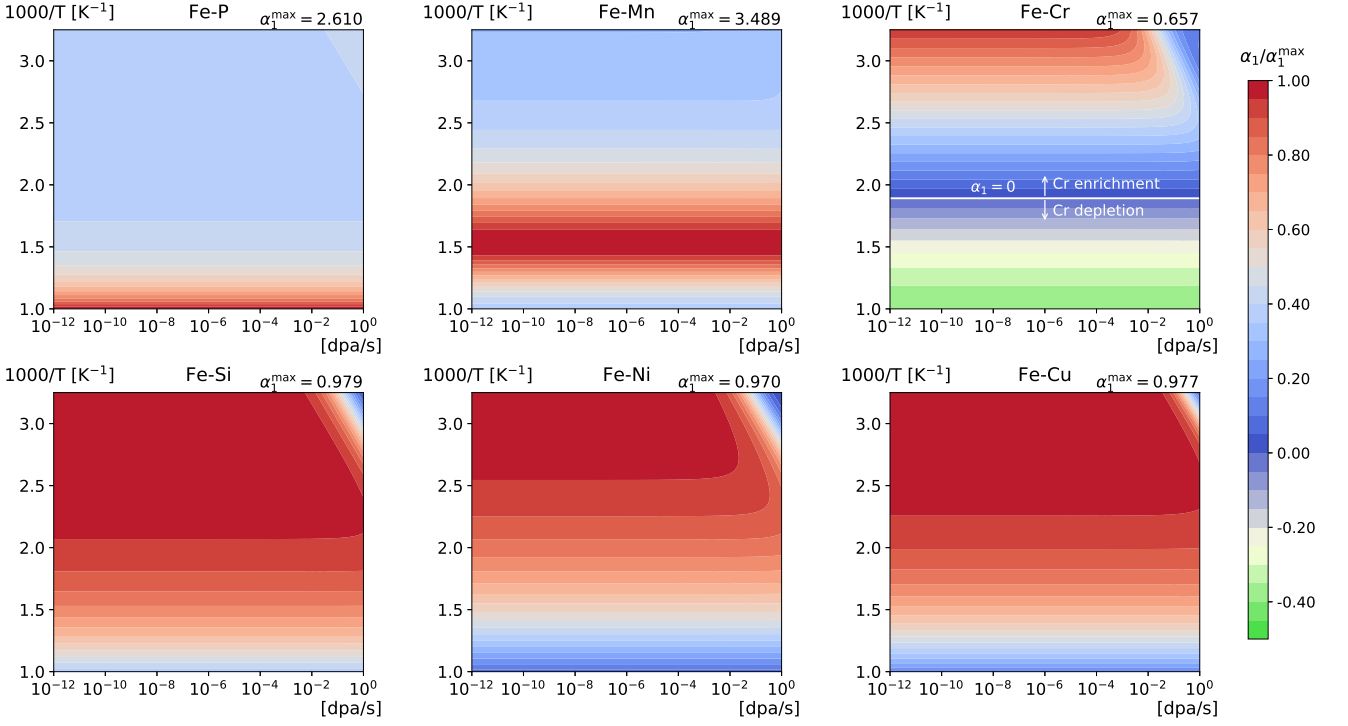}
    \caption{RIS factor $\alpha_1$ normalized by its maximum value $\alpha_1^\text{max}$ as a function of dose rate (in dpa/s) and inverse temperature (in K$^{-1}$) in several dilute binary Fe-based alloys. The solid line in Fe-Cr system corresponds to $\alpha_1=0$. The nominal solute concentration $\overline{C}_\text{B}$ is set to 1\,at.\% and the sink strength $k^2$ is set to $5\times 10^{14}\,\text{m}^{-2}$.}
    \label{fig:alpha_1_Fe-X}
\end{figure*}

\begin{figure*}
    \centering
    \includegraphics[width=1.0\linewidth]{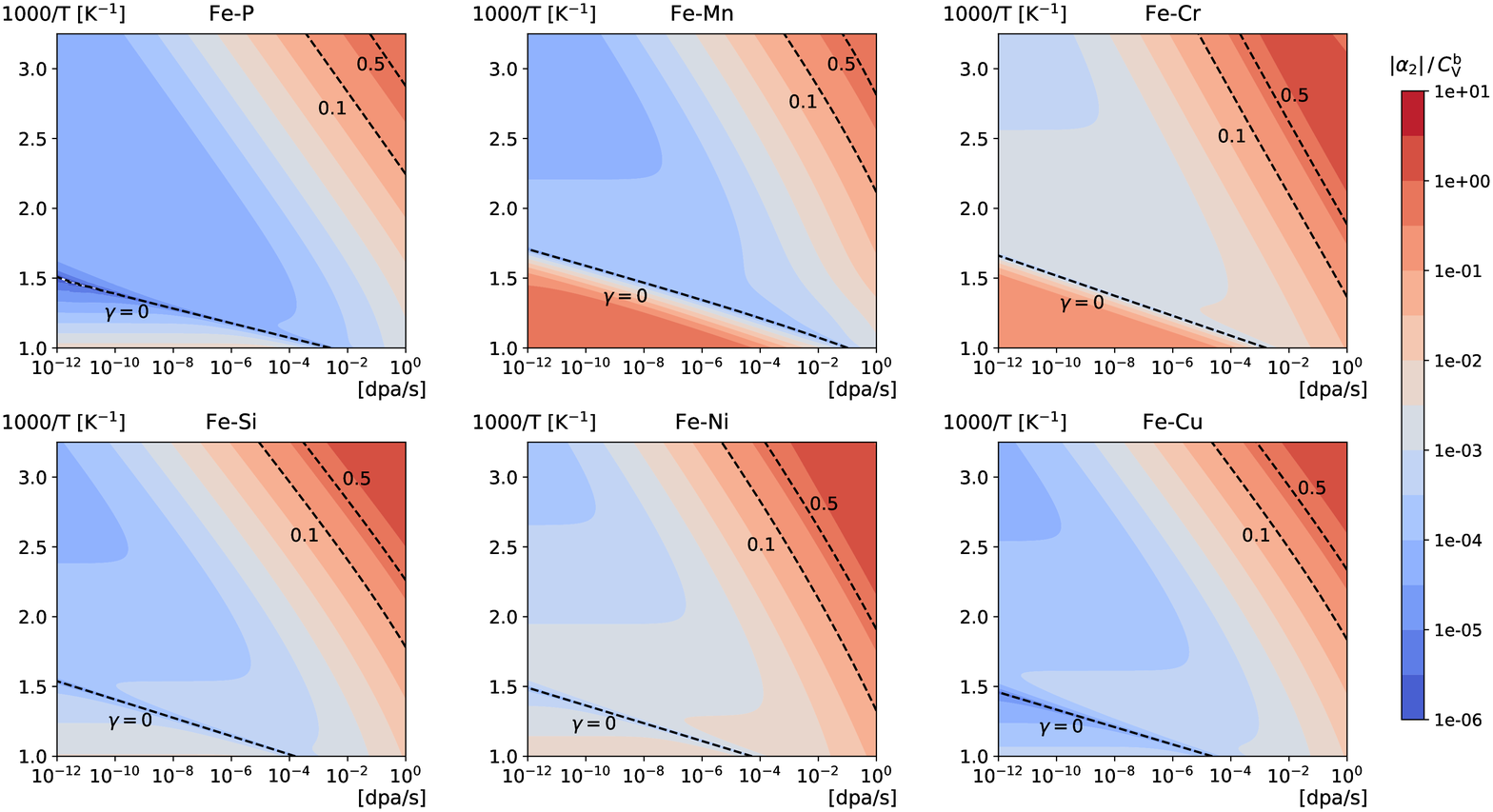}
    \caption{$|\gamma|=|\alpha_2|/C_\text{V}^\text{b}$ as a function of dose rate (in dpa/s) and inverse temperature (in K$^{-1}$) in several dilute binary Fe-based alloys. The nominal solute concentration $\overline{C}_\text{B}$ is set to 1\,at.\% and the sink strength $k^2$ is set to $5\times 10^{14}\,\text{m}^{-2}$.}
    \label{fig:alpha_2_Fe-X}
\end{figure*}

\begin{figure*}
    \centering
    \includegraphics[width=1.0\linewidth]{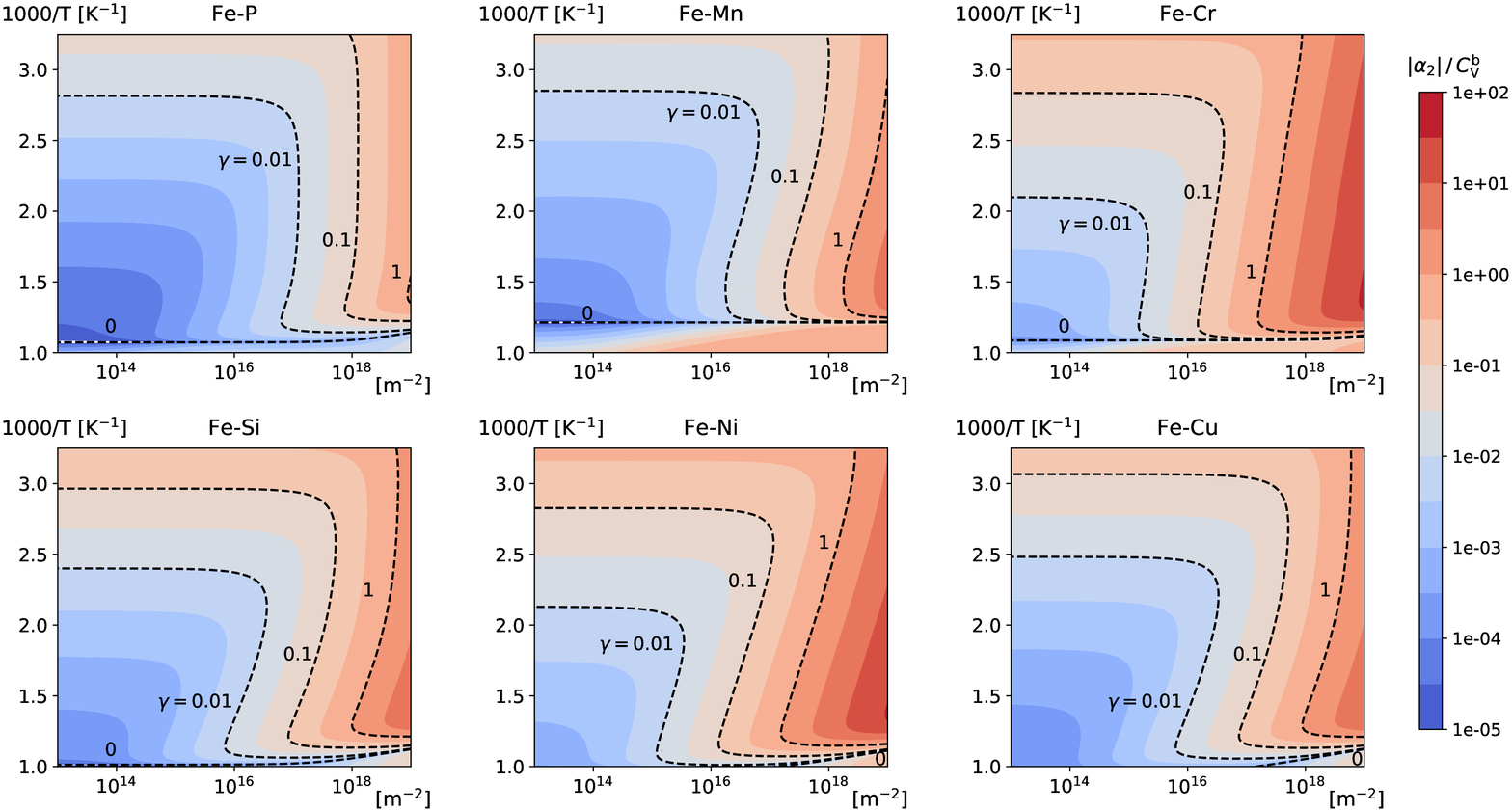}
    \caption{$|\gamma|=|\alpha_2|/C_\text{V}^\text{b}$ as a function of sink strength (in $\text{m}^{-2}$) and inverse temperature (in K$^{-1}$) in several dilute binary Fe-based alloys. The nominal solute concentration $\overline{C}_\text{B}$ is set to 1\,at.\% and $\phi$ is set to $2\times 10^{-4}$\,dpa/s.}
    \label{fig:alpha_2_Fe-X_k2}
\end{figure*}

\subsubsection{Flux coupling coefficients}
{As expected from the solute-PD binding energy database (Tab.\,\ref{tab:solute-PD_binding_Fe-X}), the overall RIS of solute atoms in Fe-Si, Fe-Ni and Fe-Cu alloys is mainly due to flux coupling mediated by the vacancy diffusion mechanism, whereas in Fe-P, Fe-Mn, and Fe-Cr alloys both the dumbbell and vacancy mechanisms contribute to the solute RIS~\cite{Messina2019}.} In addition, in Fe-Cr alloys flux coupling mediated by vacancies and SIAs have opposite sign, and the subsequent sign and amplitude of Cr RIS results from a temperature-dependent balance between the vacancy-induced depletion and the dumbbell-induced enrichment.

In Fig.\,\ref{fig:delta_Fe-X}, we plot the variation of the flux coupling coefficients with dose rate in Fe-Mn, Fe-Cr, and Fe-Cu alloys at different temperatures. For $\phi<10^{-2}$\,dpa/s, we observe that in the Fe-Cu alloy, $\delta_\text{I}$ is much smaller than $\delta_\text{V}$ because of the instability of the mixed solute-dumbbell configuration. We observe similar trends in Fe-Si and Fe-Ni alloys. For Fe-Mn, Fe-P and Fe-Cr alloys, both $\delta_\text{I}$ and $\delta_\text{V}$ have a non-negligible contribution. The above results are consistent with the flux coupling behaviors presented in Ref.\,\cite{Messina2019}, even though the FAR mechanism was therein not included.
In Ref.\,\cite{Huang2019}, we have shown  that there is an effect of FAR  whenever the jump frequencies of PDs at equilibrium are comparable with FAR frequencies within 1 to 2 orders of magnitude. Therefore, as expected, we observe an effect of FAR on $\delta_\text{V}$ at significant dose rates ($>10^{-2}$\,dpa/s) and low temperatures ($<400$\,K), a flux-temperature domain in which the FAR frequency is close to or higher than thermal jump frequencies. Note that at temperatures above $350$ K, the flux-coupling coefficient $\delta_\text{I}$ is nearly independent of dose rate below 1\,dpa/s. This is because the dumbbell-mediated jump frequencies are much higher than the FAR frequency.

\subsubsection{RIS factor \texorpdfstring{$\alpha_1$}{\alpha1}}
The RIS factor results from the balance between  flux coupling and  backward diffusion opposing the segregation of solute atoms at sinks.
Here, we consider the RIS factor $\alpha_1$, which, in the absence of FAR, corresponds to the overall RIS factor (cf.  Eq.\,\eqref{eq:alpha_1}). In Fig.\,\ref{fig:alpha_1_Fe-X}, we show the temperature-radiation flux maps of $\alpha_1$. {As expected from the solute-PD binding energies ($E_\text{b}$), $\alpha_1$ has the same behavior in Fe-Si, Fe-Ni, and Fe-Cu alloys. When $\phi<10^{-2}$\,dpa/s, the absolute value of $\alpha_1$ decreases with temperature because of the drop of the vacancy-solute pair probability {proportional to $\exp(-E_\mathrm{b}/k_\mathrm{B}T)$}~\cite{Messina2014}. In the Fe-P, Fe-Mn, and Fe-Cr systems, the variation of $\alpha_1$ with temperature is quite different.} In Fe-P, the absolute value of $\alpha_1$ increases with temperature. 
As for the Fe-Mn alloy, $|\alpha_1|$ increases up to around 650\,K. The binding energy of the Fe-Mn dumbbell is lower than that of the Fe-P dumbbell. As a consequence, above 650\,K, $|\alpha_1|$ decreases with temperature. Regarding the Fe-Cr alloy, we observe a change of sign of $\alpha_1$ around 530\,K.  

At low temperatures, roughly below 600\,K, $\alpha_1$ decreases with radiation flux. At temperatures below 300\,K and under very high radiation fluxes (above $\phi=1$\,dpa/s),  $\alpha_1$ is close to 0 because flux coupling is totally destroyed by FAR. Above 600\,K and/or below $10^{-2}$\,dpa/s, there is no effect of FAR on $\alpha_1$. Note that even though FAR may reduce the magnitude of $\alpha_1$, it does not qualitatively change the extent of the three kinetic domains, nor the sign of the RIS factor. 

\subsubsection{RIS factor \texorpdfstring{$\alpha_2$}{\alpha2}} \label{subsubsec:alpha_2}
In addition to $\alpha_1$, the RIS magnitude depends also on the RIS factor $\alpha_2$, which is directly related to the FAR mechanism (i.e., $L_\text{BB}^\text{mono}$). As stated in Sec.\,\ref{subsubsec:alpha}, the ratio $\gamma = \alpha_2/C_\text{V}^\text{b}$ indicates the extent of the FAR effect on the solute RIS. If $\gamma\gg 1$, $S_\text{B}$ is equal to 0. In Fig. \ref{fig:alpha_2_Fe-X}, we plot the $\phi$--$T$ maps of $|\gamma|=|\alpha_2|/C_\text{V}^\text{b}$. Dashed lines represent level lines of $\gamma$. We observe that over most flux-temperature conditions, $\gamma$ is smaller than $0.1$. It is close to or larger than 1 at high  $\phi$ and low $T$, and increases with dose rate. Furthermore, it decreases with temperature, and becomes negative above a threshold temperature linearly increasing with radiation flux. The domain where $\gamma<0$ coincides with the thermal domain of Fig.\,\ref{fig:CV_Fe-X}. At the limit of this domain, $C_\text{V}^\text{b}$ is close to the equilibrium vacancy concentration $C_\text{V}^\text{eq}$. 

In order to highlight the effect of sink strength on $\gamma$, we plot the $k^2$-$T$ maps of $|\gamma|$ in Fig.\,\ref{fig:alpha_2_Fe-X_k2}. Since $\gamma$ increases with $k^2$, the FAR effect on RIS should be significant at large values of $k^2$. 

Note that the variation of $\gamma$ with $T$, $\phi$, and $k^2$ is alloy-specific. For a given set of parameters, the value of $\gamma$ is relatively high in Fe-Ni and Fe-Cr, small in Fe-P and Fe-Mn, and in an intermediate range in Fe-Si and Fe-Cu alloys.

\subsection{Radiation induced segregation of PDs}

\begin{figure}
    \centering
    \includegraphics[width=1.0\linewidth]{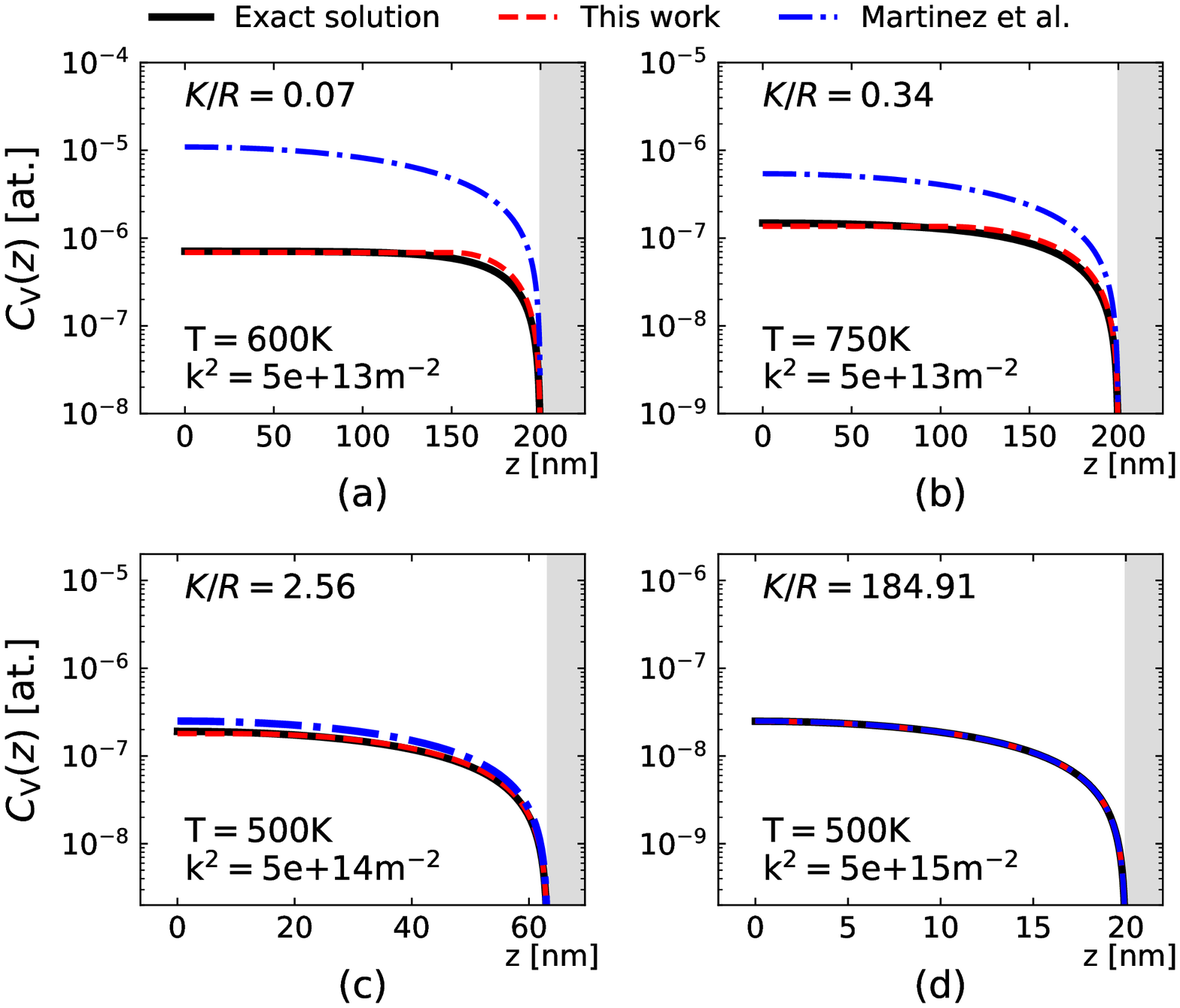}
    \caption{Concentration profiles of vacancies under irradiation. The solid lines are the exact solutions of Eq.\,\eqref{eq:CV_diffusion_2}. The dashed and dash-dotted lines are the analytical approximations of Eq.\,\eqref{eq:CV_diffusion_2}, obtained from Eq.\,\eqref{eq:CV_two_regimes} (this work) and Eq.\,\eqref{eq:CV_prl} (cf. Ref.\,\cite{Martinez2018}), respectively. The shaded area indicates the sink. The results for (a) and (b) are respectively given for $T=$600\,K and 750\,K, with $\phi=10^{-4}$\,dpa/s and $k^2=5\times 10^{13}\,\text{m}^{-2}$ (i.e., $h=400$\,nm). The results for (c) and (d) are respectively given for $k^2=5\times 10^{14}\,\text{m}^{-2}$ and $5\times 10^{15}\,\text{m}^{-2}$, with $\phi=10^{-6}$\,dpa/s and $T=$500\,K.}
    \label{fig:CV_Fe-X_T_k2}
\end{figure}

{In order to investigate the effect of PD recombination on the RIS profiles, we compare the profiles given by two different methods: (i)\,the analytical approximation proposed in this work (Eq.\,\eqref{eq:CV_two_regimes} for $C_\text{V}(z)$), and (ii)\,the one proposed in~Ref.\,\cite{Martinez2018} where the recombination rate is set to zero.  

The concentration profiles of vacancies at different temperatures and sink strengths are plotted in Fig.\,\ref{fig:CV_Fe-X_T_k2}. In order to assess the accuracy of the analytical approximations, we plot as well the reference profile obtained from the exact solution of Eq.\,\eqref{eq:CV_diffusion_2} computed by a finite-difference method. We observe that the concentration profiles obtained from the {present} analytical approach are in good agreement with the reference profiles. 

When recombination reactions are neglected, the vacancy concentration along the RIS profile is overestimated, especially at low temperatures (e.g., 600\,K) and small sink strength (e.g., $5\times 10^{13}\,\text{m}^{-2}$), because the ratio $K/R$ is relatively large. Therefore, the recombination effect is not negligible (see Sec.\,\ref{subsubsec:RIS_PD}). }

\begin{figure*}
    \centering
    \includegraphics[width=1.0\linewidth]{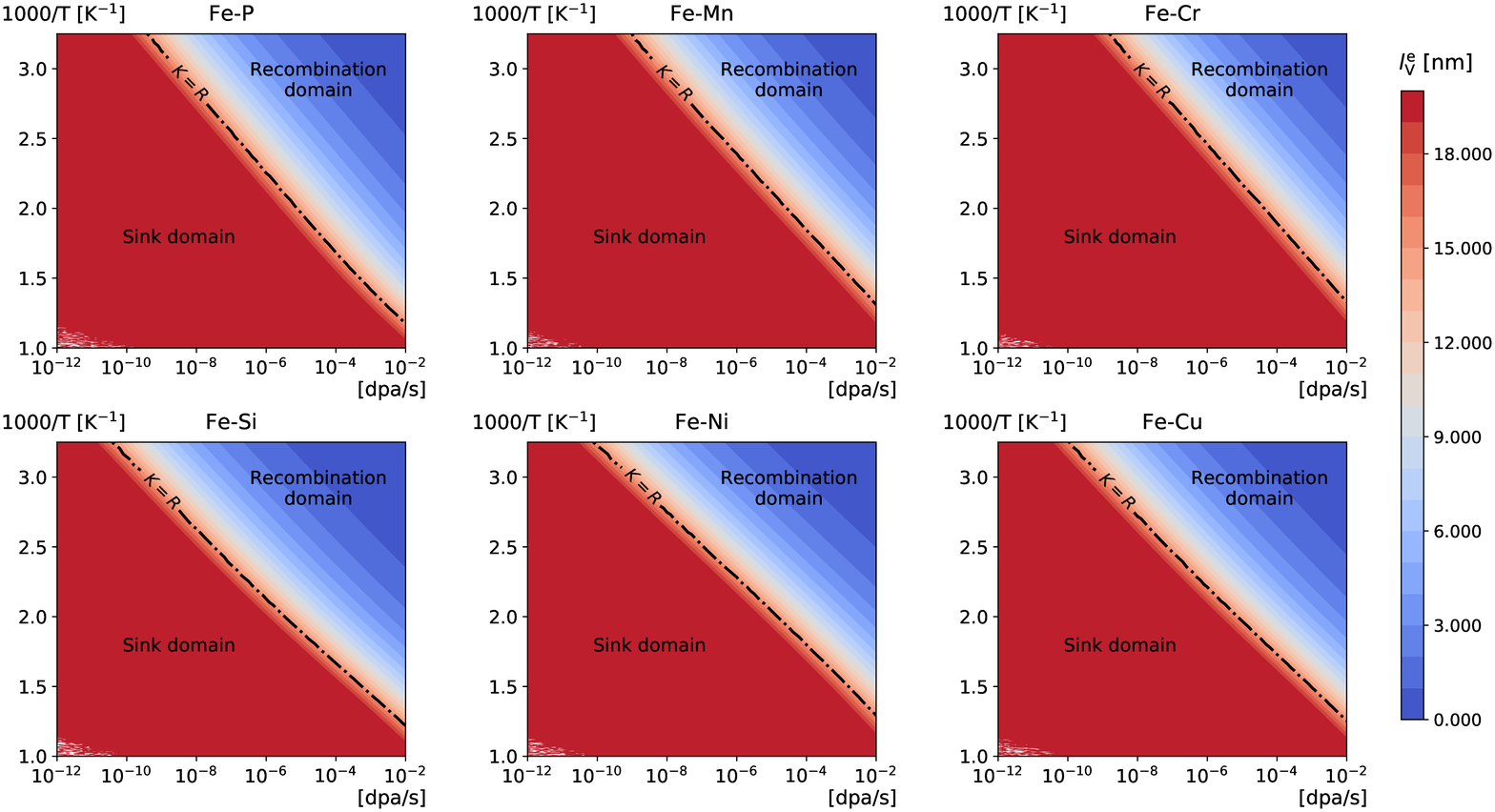}
    \caption{The effective width of RIS profiles of vacancies as a function of dose rate (in dpa/s) and inverse temperature (in K$^{-1}$) for several dilute binary Fe-based alloys. The nominal solute concentration $\overline{C}_\text{B}$ is set to 1\,at.\% and the sink strength $k^2$ is set to $5\times10^{14}\,\text{m}^{-2}$. The corresponding distance between planar sinks is $h=126$ nm.}
    \label{fig:leV_Fe-X}
\end{figure*}
In order to investigate the shape of the vacancy RIS profile at different irradiation conditions, we define an effective width $l_\text{V}^\text{e}$ of the vacancy concentration profile as follows:
\begin{equation} \label{eq:leV}
    l_\text{V}^\text{e} = \sqrt{\frac{\int_0^{h/2}\left[(h/2)-z \right]^2 \left[ C_\text{V}(z)-C_\text{V}(0) \right]\text{d}z}{\int_0^{h/2} [C_\text{V}(z)-C_\text{V}(0)] \text{d}z}}.
\end{equation}
{This parameter represents the average distance between a vacancy and a PD sink. It is also related to the width of the vacancy-depleted zone near sinks~\cite{Foreman1972}. In Fig.\,\ref{fig:leV_Fe-X}, we plot the maps of $l_\text{V}^\text{e}$ as a function of the inverse temperature and dose rate. In the sink domain, the width of the vacancy profile does not change with the irradiation conditions. According to the analytical solution of Eq.\,\eqref{eq:leV}, $l_\text{V}^\text{e}$ increases with $h$. Therefore, the smaller the sink density, the larger the distance between sinks, and the wider the vacancy-depleted zone. In the recombination domain, $l_\text{V}^\text{e}$ decreases with dose rate, while it increases with temperature. }

\begin{figure*}
    \centering
    \includegraphics[width=1.0\linewidth]{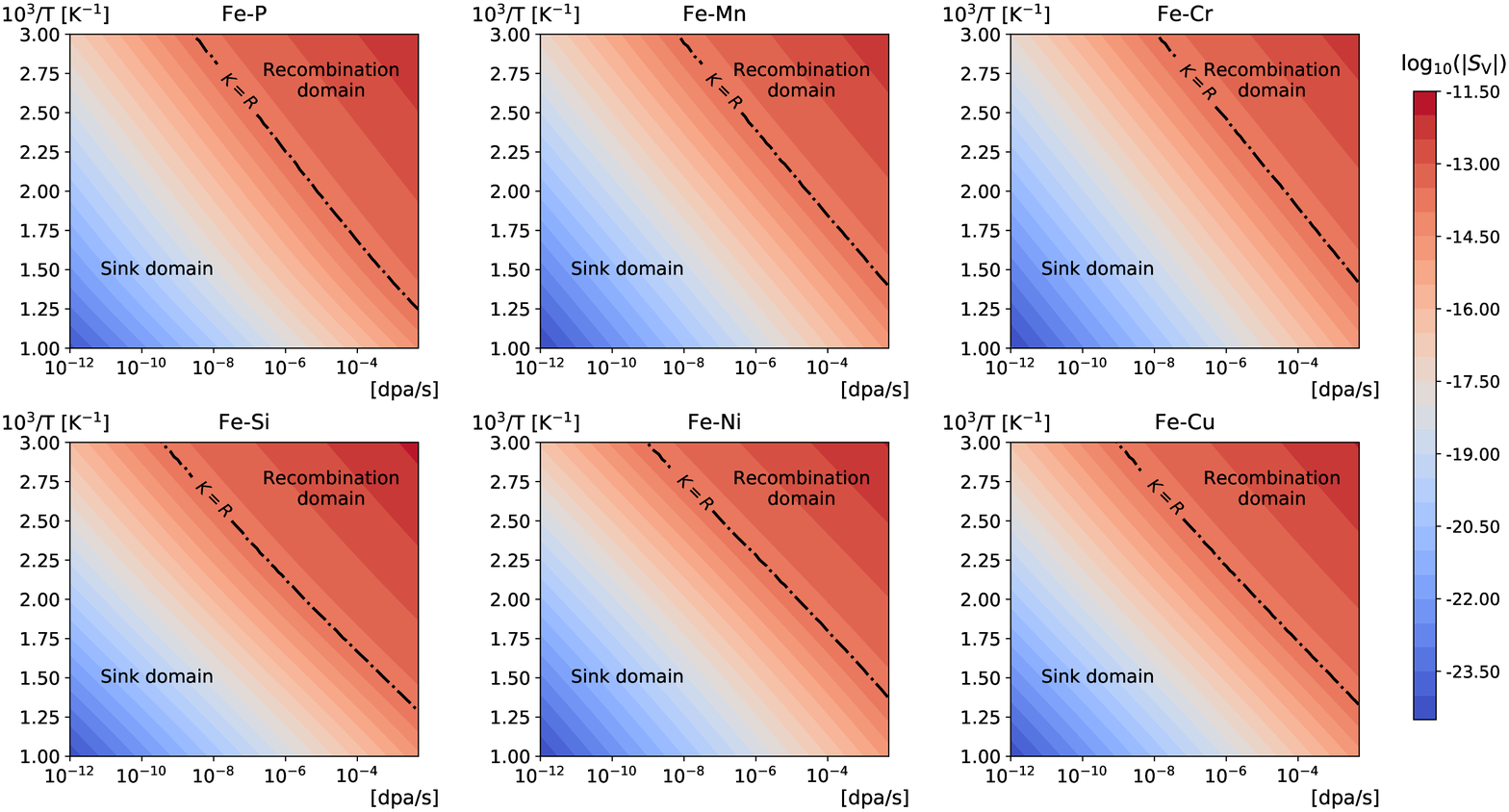}
    \caption{Total amount of segregated vacancies $S_\text{V}$ as a function of dose rate (in dpa/s) and inverse temperature (in K$^{-1}$) for several dilute binary Fe-based alloys. The nominal solute concentration $\overline{C}_\text{B}$ is set to 1\,at.\% and the sink strength $k^2$ is set to $5\times10^{14}\,\text{m}^{-2}$.}
    \label{fig:SV_Fe-X}
\end{figure*}

\begin{figure*}
    \centering
    \includegraphics[width=1.0\linewidth]{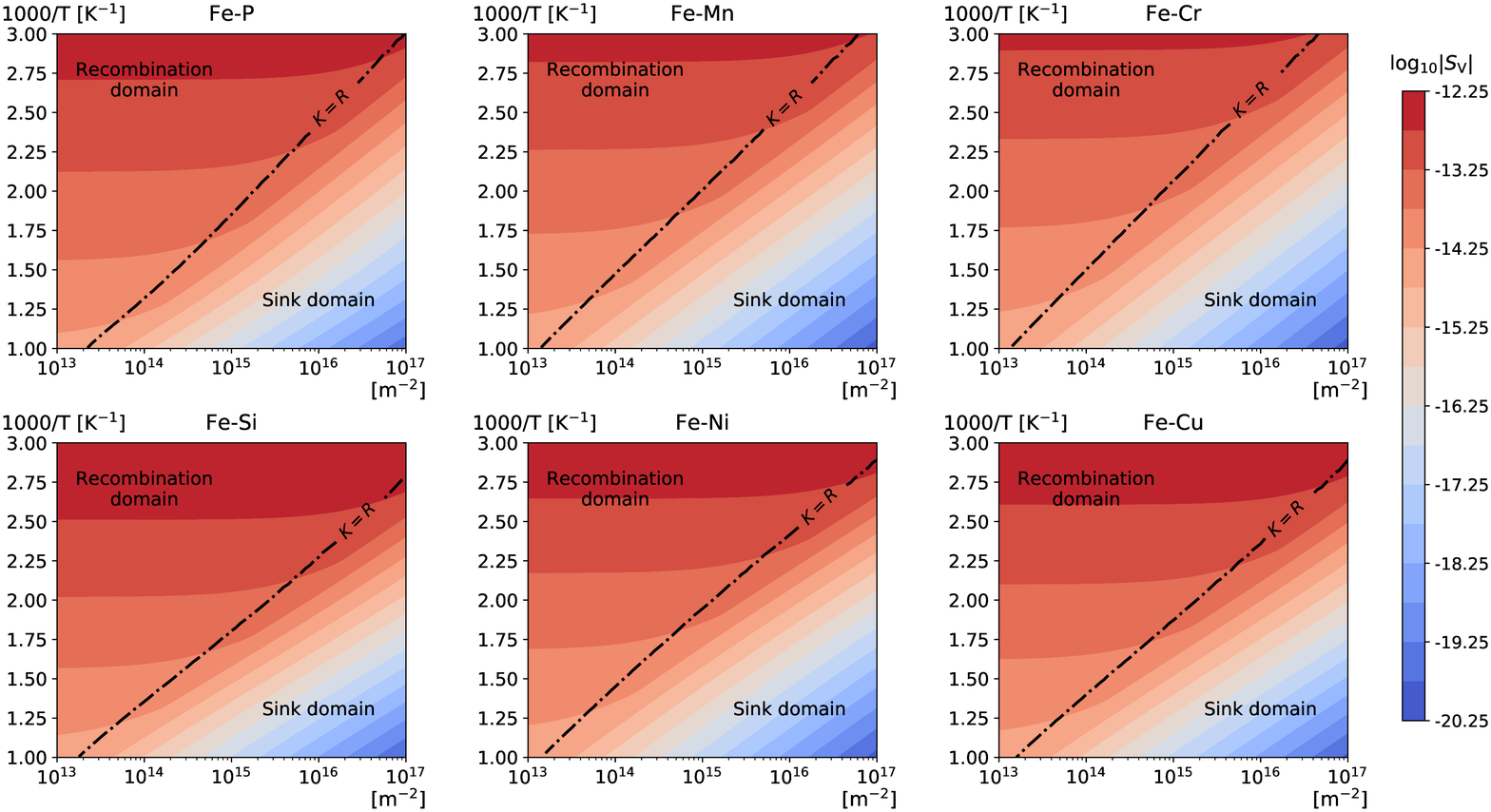}
    \caption{Total amount of segregated vacancies $S_\text{V}$ as a function of sink strength (in m$^{-2}$) and inverse temperature (in K$^{-1}$) for several dilute binary Fe-based alloys. The nominal solute concentration $\overline{C}_\text{B}$ is set to 1\,at.\% and the dose rate $\phi$ is set to $2\times10^{-4}$\,dpa/s.}
    \label{fig:SV_Fe-X_k2}
\end{figure*}

Furthermore, we investigate the effect of temperature, dose rate, and sink strength on the segregation amount of vacancies. Note that, using Eq.\,\eqref{eq:SV_elimination_dominant} and Eq.\,\eqref{eq:SV_recombination_dominant}, $\text{log}\,|S_\text{V}|$ is given by
\begin{equation} \label{eq:log_SV}
    \text{log}\,|S_\text{V}|=
    \begin{cases}
        \text{log}\,\phi-\text{log}\,D_\text{V}-\frac{3}{2}\text{log}\,k^2 + K_2, & \text{$K\gg R$}; \\
        \frac{1}{4}\text{log}\,\phi-\frac{1}{4}\text{log}\,D_\text{V} + K_3, & \text{$K\ll R$},
    \end{cases}
\end{equation}
with $K_2=\text{log}\left({\sqrt{2}}/{3}\right)$ and $K_3=\text{log}\left[\left({\Omega}/{\pi r_\text{c}}\right)^{3/4}/6\right]$.

Fig.\,\ref{fig:SV_Fe-X} shows the maps of $\text{log}\,|S_\text{V}|$ near the interface (given by Eq.\,\eqref{eq:SV}) as a function of inverse temperature and dose rate. The maps are divided into two domains corresponding to the two limit cases of Eq.\,\eqref{eq:log_SV}. The first kinetic domain is the one dominated by recombination reactions  ($K<R$), and the second one is the sink domain ($K>R$). In both regimes, $\text{log}\,|S_\text{V}|$ increases linearly with $\text{log}\,\phi$ and $1/T$, but the slopes are different. 

Fig.\,\ref{fig:SV_Fe-X_k2} shows the temperature-sink strength maps of the same quantity, divided as well into recombination and sink domains. $\text{log}\,|S_\text{V}|$ decreases linearly with $\text{log}\,k^2$ in the sink domain, whereas it is nearly $k^2$-independent in the recombination domain. 
The variations of $S_\text{V}$ with $\phi$ and $k^2$ are similar in all investigated Fe-based binary alloys, whereas the variations with $1/T$ are a bit more varied because the average vacancy diffusion coefficient $D_\text{V}$ is alloy-specific.

\subsection{Radiation induced segregation of solute atoms}
\begin{figure*}
    \centering
    \includegraphics[width=1.0\linewidth]{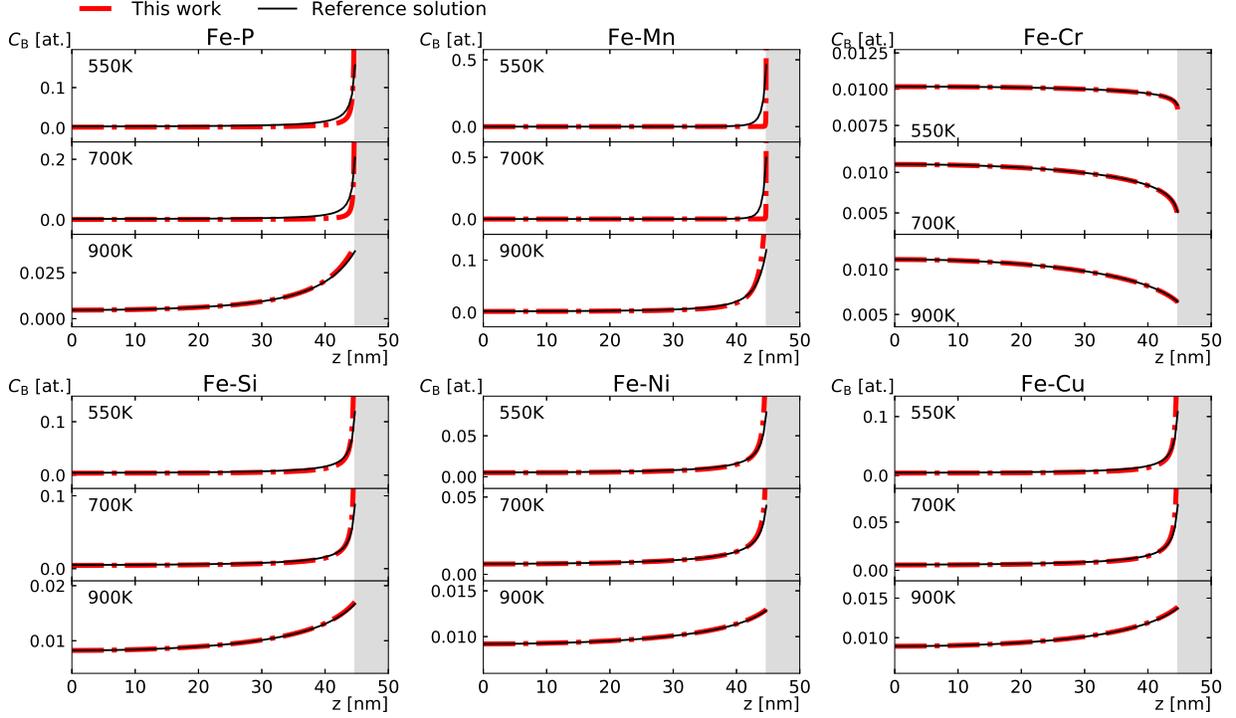}
    \caption{Solute RIS profiles obtained by our analytical models in several dilute Fe-based alloys. The profiles obtained by the exact solution of Eq.\,\eqref{eq:grad_CB_CV} are also plotted as references. The nominal solute concentration $\overline{C}_\text{B}$ is set to 1\,at.\% and the sink strength $k^2$ is set to $10^{15}\,\text{m}^{-2}$. The corresponding distance between planar sinks is $h=86$ nm.}
    \label{fig:CB_Fe-X_profiles}
\end{figure*}
\begin{figure*}
    \centering
    \includegraphics[width=1.0\linewidth]{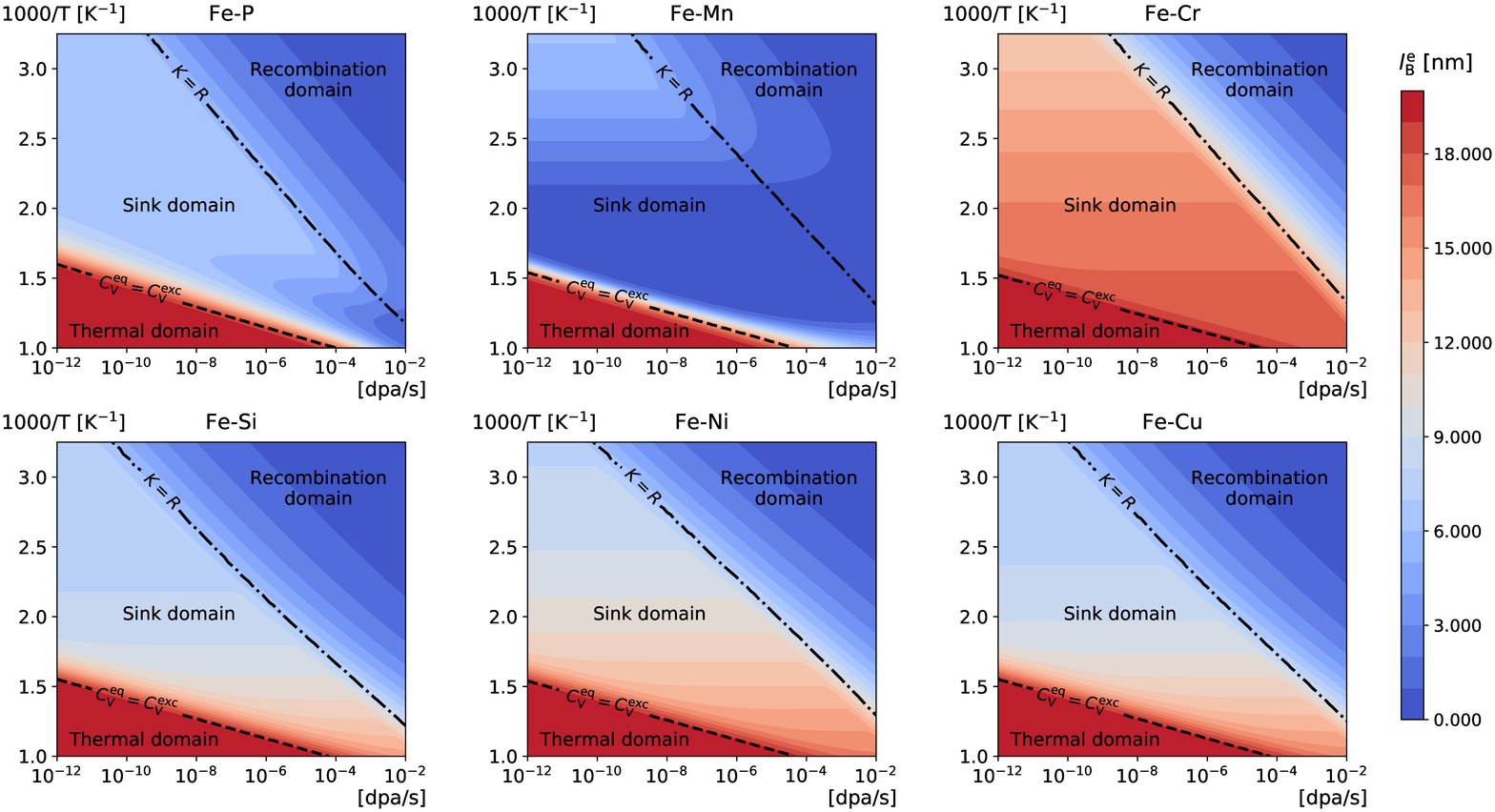}
    \caption{The effective width of RIS profiles of solute atoms as a function of dose rate (in dpa/s) and inverse temperature (in K$^{-1}$) for several dilute binary Fe-based alloys. The nominal solute concentration $\overline{C}_\text{B}$ is set to 1\,at.\% and the sink strength $k^2$ is set to $5\times10^{14}\,\text{m}^{-2}$. The corresponding distance between planar sinks is $h=126$ nm. The dose rate is set to $10^{-5}$\,dpa/s.}
    \label{fig:leB_Fe-X}
\end{figure*}

\begin{figure*}
    \centering
    \includegraphics[width=1.0\linewidth]{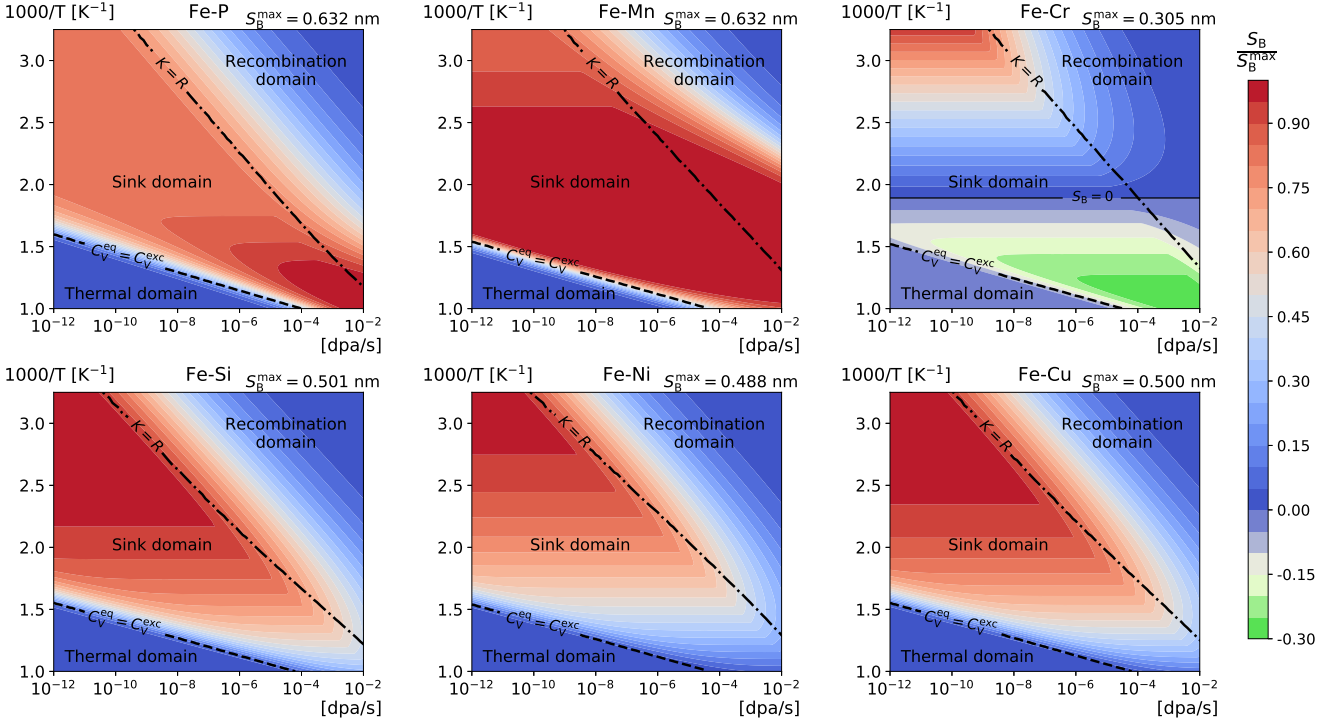}
    \caption{Total amount of segregated solute atoms $S_\text{B}$ normalized by its maximum over all considered irradiation conditions $S_\text{B}^\text{max}$ as a function of dose rate (in dpa/s) and inverse temperature (in K$^{-1}$) for several dilute binary Fe-based alloys. The nominal solute concentration $\overline{C}_\text{B}$ is set to 1\,at.\% and the sink strength $k^2$ is set to $5\times 10^{14}\,\text{m}^{-2}$.}
    \label{fig:SB_Fe-X}
\end{figure*}

\begin{figure*}
    \centering
    \includegraphics[width=1.0\linewidth]{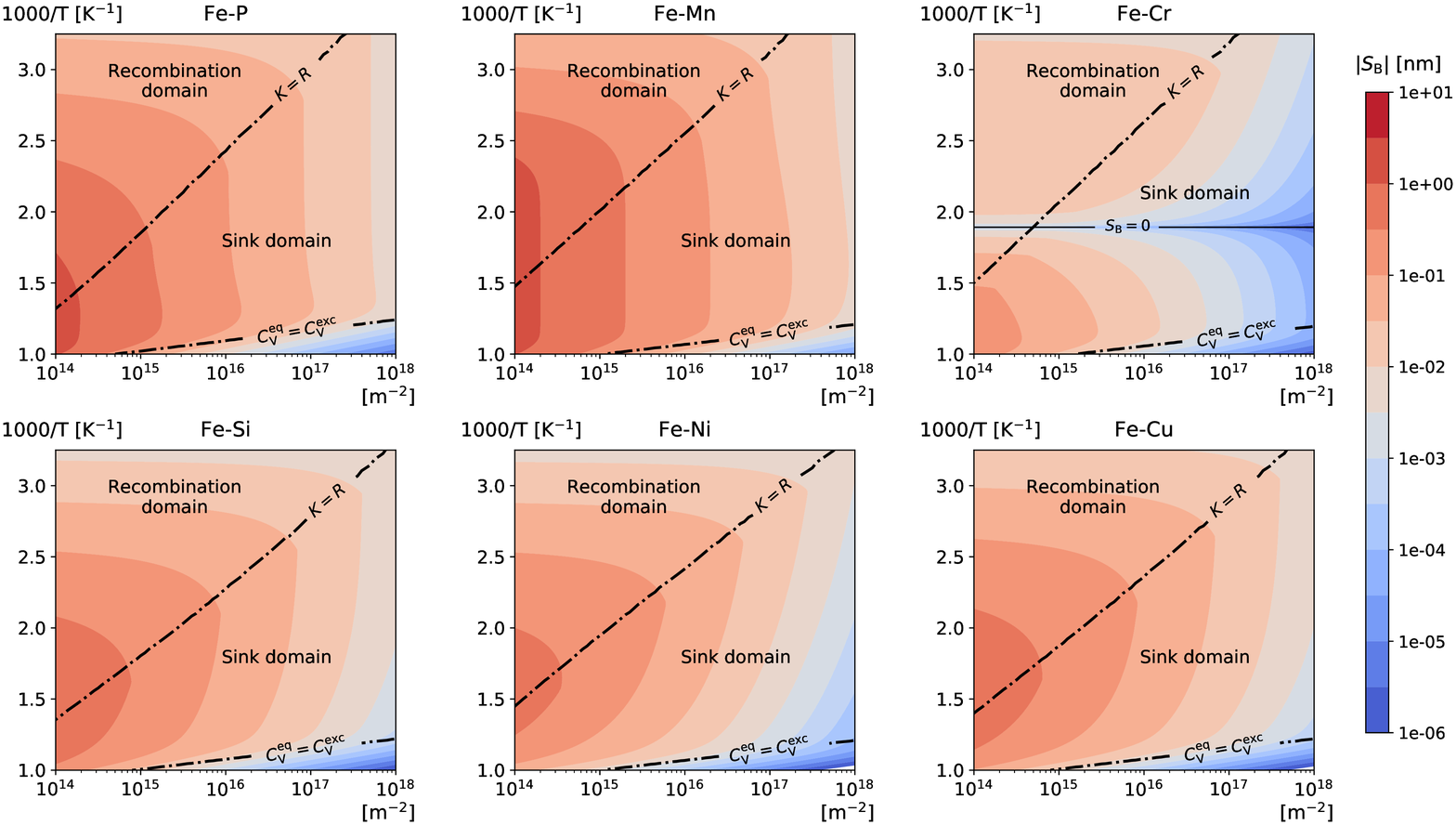}
    \caption{Total amount of segregated solute atoms $S_\text{B}$ as a function of sink strength (in m$^{-2}$) and inverse temperature (in K$^{-1}$) for several dilute binary Fe-based alloys. The nominal solute concentration $\overline{C}_\text{B}$ is set to 1\,at.\% and the dose rate $\phi=2\times 10^{-4}$\,dpa/s.}
    \label{fig:SB_Fe-X_k2}
\end{figure*}

\begin{figure*}
    \centering
    \includegraphics[width=1.0\linewidth]{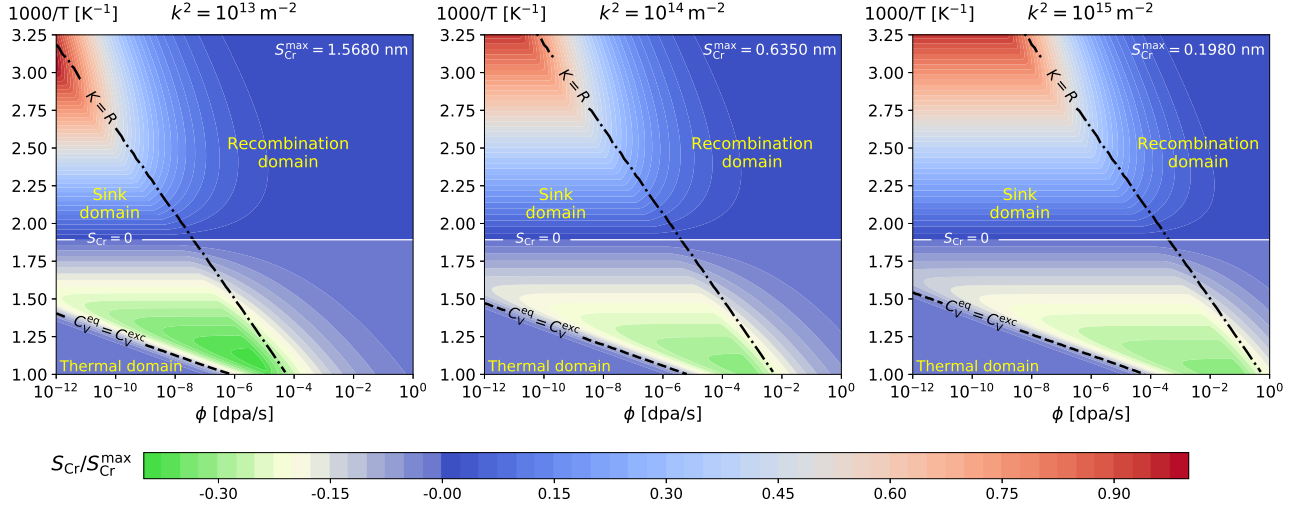}
    \caption{Total amount of segregated Cr atoms $S_\text{Cr}$ as a function of dose rate (in dpa/s) and inverse temperature (in K$^{-1}$) computed for different sink strengths: $10^{13}\,\text{m}^{-2}$, $10^{14}\,\text{m}^{-2}$ and $10^{15}\,\text{m}^{-2}$. The results are normalized by the maximum segregation amount $S_\text{Cr}^\text{max}$. The nominal solute concentration $\overline{C}_\text{B}$ is set to 1\,at.\%.}
    \label{fig:SB_Cr_k2}
\end{figure*}

\begin{figure*}
    \centering
    \includegraphics[width=1.0\linewidth]{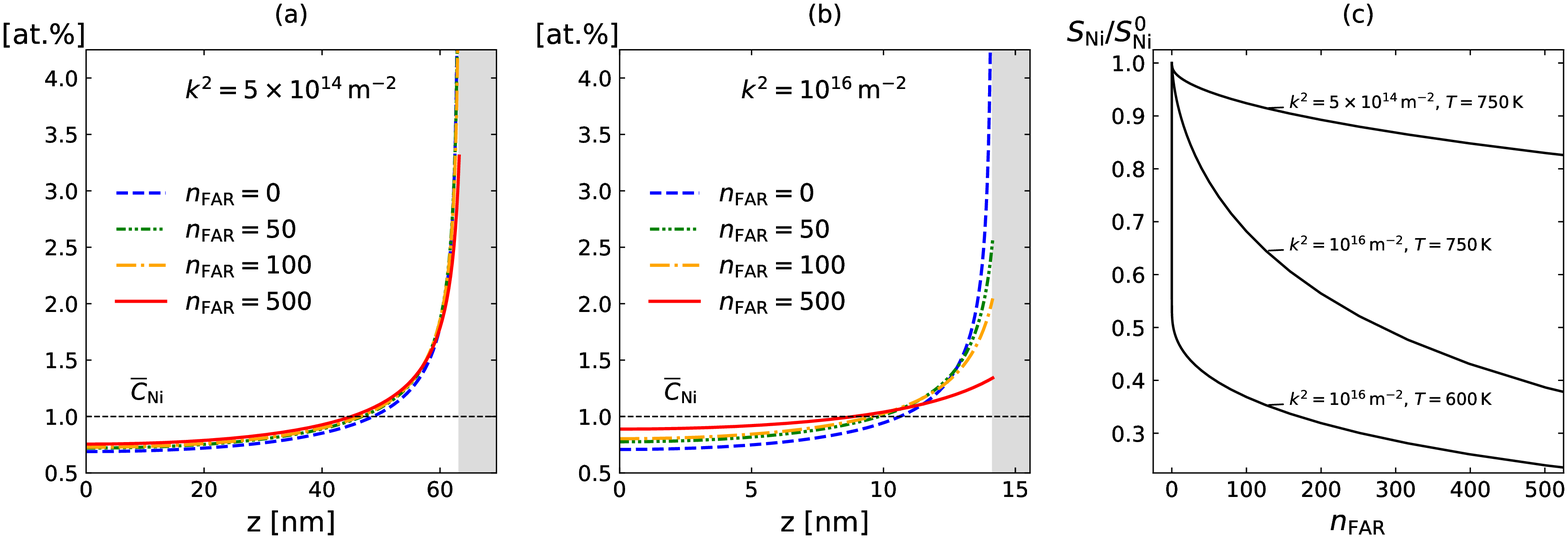}
    \caption{Ni segregation profile $C_\text{Ni}(z)$ near an interface (indicated by the shaded area) computed for (a)\,$k^2=5\times10^{14}$\,m$^{-2}$ and $T=750$\,K; (b)\,$k^2=10^{16}$\,m$^{-2}$ and $T=750$\,K; and (c) Ni segregation amount with different FAR intensities at $T=600$ and $700\,$K. $S_\text{Ni}^0$ is the segregation amount obtained with $n_\text{FAR}=0$. The nominal solute concentration $\overline{C}_\text{Ni}$ is set to 1\,at.\% and the dose rate $\phi$ is set to $10^{-4}$\,dpa/s.}
    \label{fig:CB_Ni_FAR}
\end{figure*}

In the analytical model, $\alpha_1$ and $\alpha_2$ are assumed to be independent of the local atomic concentration $C_\text{B}$, although it is shown in Sec.\,\ref{subsubsec:alpha} that this is not true in general. For validation, we compare the RIS profiles given by the analytical model (Eq.\,\eqref{eq:CB_two_regimes}) and a reference profile obtained by the numerical integration of Eq.\,\eqref{eq:grad_CB_CV}. The variation of $\alpha_1$ and $\alpha_2$ with $C_\text{B}$ is entirely accounted for in the reference profile. The solute RIS profiles at different temperatures are plotted in Fig.\,\ref{fig:CB_Fe-X_profiles}. In Fe-Cr, Fe-Si, Fe-Ni, and Fe-Cu alloys, the analytical profiles are in very good agreement with the numerical results at every temperature. In Fe-P and Fe-Mn alloys, we observe slightly different profile shapes between the analytical and reference results at 550 and 700\,K. However, we systematically obtain a very good agreement on the solute bulk concentration and the solute segregation amount.

As for the vacancy concentration profile, we introduce an effective width of the concentration profile of solute atoms ($l_\text{B}^\text{e}$) to characterize the shape of the solute RIS profile. Its definition is similar to that of the vacancies (Eq.\,\eqref{eq:leV}), where subscript V is replaced by B. Fig.\,\ref{fig:leB_Fe-X} shows the $\phi$--$T$ maps of $l_\text{B}^\text{e}$. This quantity is large and almost uniform in the thermal domain. In the recombination domain, $l_\text{B}^\text{e}$ decreases with dose rate and increases with temperature. These trends are very similar to the ones observed for $l_\text{V}^\text{e}$. As a result, we expect the RIS profiles of vacancies and solute atoms to have almost the same width in thermal and recombination domains. In these domains, we could rely on the measured width of the solute RIS profiles to obtain information on the vacancy RIS profile, and subsequently on the PDs sink strength. On the other hand, in the sink domain, the larger the solute RIS amount, $S_\text{B}$, the smaller the width of the RIS profile, $l_\text{B}^\text{e}$. Moreover, $l_\text{B}^\text{e}$ is smaller than $l_\text{V}^\text{e}$ in this domain, especially in Fe-P and Fe-Mn alloys where the tendency to positive RIS is significant.

Furthermore, Figs.\,\ref{fig:SB_Fe-X} and \ref{fig:SB_Fe-X_k2} show the temperature-radiation flux-sink strength maps of the solute RIS amount given by Eq.\,\eqref{eq:SB_two_regimes_2}. As shown in Fig.\,\ref{fig:DB_Fe-X}, the flux-temperature ($\phi$--$T$) domains of $S_\text{B}$ are mainly determined by PD kinetics. $S_\text{B}$ is significant in the sink domain, whereas it is relatively small in the recombination domain because, after the SIA-V recombination, only a few PDs are left for the long-distance solute diffusion towards the sinks. In the thermal domain, since the amount of excess PDs is very small, so is the net flux of PDs towards sinks, which leads to small $S_\text{V}$ and $S_\text{B}$.

In Fig.\,\ref{fig:SB_Fe-X}, we show the maps of the solute RIS amount $S_\text{B}$ as a function of $T$ and $\phi$ at fixed sink strength $k^2=5\times10^{14}$\,m$^{-2}$. As expected, $S_\text{B}$ does not depend on dose rate in the sink domain (cf. Eq.\,\eqref{eq:SB_elimination_dominant}), and $S_\text{B}$ decreases with dose rate in the recombination domain (cf. Eq.\,\eqref{eq:SB_recombination_dominant}). 
In Fig.\,\ref{fig:SB_Fe-X_k2}, we show the maps of $S_\text{B}$ as a function of $k^2$ and $T$ at fixed radiation flux $\phi=2\times10^{-4}$\,dpa/s. In the domain of PD elimination ($K>R$), $S_\text{B}$ decreases with $k^2$, as expected from Eq.\,\eqref{eq:SB_elimination_dominant}. 

We take the Fe-Cr alloy as an example to illustrate the effect of sink strength on the extent of the kinetic domains and the maximum of solute RIS. Fig.\,\ref{fig:SB_Cr_k2} shows the Cr RIS amount maps at different sink strength $k^2$. Note that the sink domain grows wider with increasing $k^2$. However, the maximum RIS amount, $S_\text{Cr}^{\text{max}}$, decreases with sink strength. Therefore, an increase of sink concentration or strength decreases the solute RIS at each sink. We obtain the same trends for the other Fe-based alloys (not represented). 

The variation of $S_\text{B}$ with $T$, $\phi$ and $k^2$ strongly depends on the chemical nature of solute atoms because $\alpha_1$ and $\alpha_2$ are alloy-specific (see Fig.\,\ref{fig:alpha_1_Fe-X} and Fig.\,\ref{fig:alpha_2_Fe-X}). {As expected from the DFT-based data of the solute-PD binding energies (\ref{tab:solute-PD_binding_Fe-X}), the main trends are similar in Fe-Si, Fe-Ni, and Fe-Cu alloys, but very different in Fe-P, Fe-Mn, and Fe-Cr alloys.} In Fe-Si, Fi-Ni, and Fe-Cu alloys, the highest solute enrichment tendency is at low temperatures (about 400\,K) and dose rate (about $10^{-12}$\,dpa/s), whereas the highest solute enrichment tendency in Fe-P is at high temperatures ($>1000$\,K) and dose rates (around $10^{-3}$\,dpa/s). As for Mn solutes, the peak of RIS occurs at intermediate temperatures (about 650\,K) and dose rate (from $10^{-10}$ to $10^{-6}$\,dpa/s). For Cr solutes, the peak of positive RIS occurs at low temperatures (about 300\,K) and dose rates (about $10^{-11}$\,dpa/s), whereas the peak of negative RIS occurs at high temperatures ($>800$\,K) and dose rates ($>10^{-6}$\,dpa/s). 

As stated in Sec.\,\ref{subsubsec:alpha_2}, the FAR effect on the solute RIS should be significant at large $k^2$ because, in this case, $\gamma$ is close to or larger than $1$. We take the Fe-Ni alloy as an example to investigate the FAR effect on the solute RIS profile. In order to identify the FAR effect, we calculate and compare the $C_\text{Ni}$ profiles and segregation amounts $S_\text{Ni}$ at different FAR intensities, which are characterized by the values of $n_\text{FAR}$ (i.e., the number of FAR per dpa). Note that $n_\text{FAR}=0$ indicates that there is no FAR in the displacement cascade. The concentration profiles are plotted in Fig.\,\ref{fig:CB_Ni_FAR} (a) and (b) with two different sink strengths and $n_\text{FAR}$ equal to $0$, $50$, $100$, and $500$. At $k^2=5\times10^{14}$\,m$^{-2}$, the segregation profiles are practically insensitive to FAR effects, because $\gamma<0.01$ and FAR can be neglected. However, at $k^2= 10^{16}$\,m$^{-2}$, the RIS profiles strongly depend on the FAR intensity $n_\text{FAR}$. The concentration of solute atoms at the interface decreases with $n_\text{FAR}$. Moreover, the amount of segregated Ni atoms also decreases with $n_\text{FAR}$ (Fig.\,\ref{fig:CB_Ni_FAR}\,(c)). At $k^2=10^{16}$\,m$^{-2}$ and $n_\text{FAR}=500$, $S_\text{Ni}$ is less than about half the one without FAR (i.e., $n_\text{FAR}=0$) at both investigated temperatures (600 and 750\,K). We observe similar tendencies in the other five Fe-based dilute alloys. {This FAR effect is significant close to the interface. In this region, the vacancy concentration is low, and the thermally-activated backward diffusion of solutes is limited. In this case, FAR is the major mechanism for backward diffusion. Therefore, RIS models ignoring FAR events overestimate the RIS tendencies in Fe-based alloys, especially at large sink strengths.} 

\section{{Discussion: dose rate compensation by a temperature shift}}

\begin{table*}
\caption{\label{tab:temperature-shift_case_definition}
Definition of cases in which quantitative criteria of temperature shift $\Delta T$ can be proposed.}
\begin{ruledtabular}
{\begin{tabular}{p{0.75cm}<{\centering} p{4cm} p{3.25cm} p{8.5cm} }
Cases & Assumptions & Criteria for $S_\text{V}$ & Criteria for $S_\text{B}$ \\
\hline
\\
(i) & $k^2$ independent of $\phi$ and $T$ & 
        Invariant $\phi/D_\text{V}$ 
 &  
        \tabitem $K\ll R$\footnotemark[1]: invariant $\frac{I_{\alpha_1}-1}{I_{\alpha_1}-1+\frac{h}{2}\left(\frac{\Omega}{\pi r_\text{c}  }\right)^{-1/4}\left(\frac{\phi}{D_\text{V}}\right)^{1/4}}$ 
      \\
   &  &  & \tabitem $K\gg R$\footnotemark[1]: no temperature shift is needed ($\Delta T=0$) \\
   &  &  & \tabitem $K\simeq R$: use of $\phi$--$T$ maps (e.g., Tab.\,\ref{tab:temperature-shift_Fe-X}) \\
\\
\hline
\\
(ii) & $K\ll R$\footnotemark[1]   & Invariant $\phi/D_\text{V}$ & Invariant $\frac{I_{\alpha_1}-1}{I_{\alpha_1}-1+\frac{h}{2}\left(\frac{\Omega}{\pi r_\text{c}  }\right)^{-1/4}\left(\frac{\phi}{D_\text{V}}\right)^{1/4}}$  \\
\\
\hline
\\
(iii) & $K\gg R$\footnotemark[1]  & Invariant $\frac{1}{\left( k^2 \right)^{3/2}}\left(\frac{\phi}{D_\text{V}}\right)$ & \tabitem If $\Delta T$ is sufficiently small such that the variation of $\alpha_1$ is negligible (e.g., within about $\pm50$\,K from Fig.\,\ref{fig:alpha_1_Fe-X}): invariant $k^2$ \\
& & & \tabitem Else: invariant $\frac{h}{2}\, \frac{I_{\alpha_1}-1}{I_{\alpha_1}-1+\frac{h}{2}\sqrt{\frac{k^2}{2}}}$ \\
\\
\hline
\\
(iv) & $k^2$($\phi$, $T$) is given & \multicolumn{2}{p{12cm}}{We use our models to calculate $S_\text{V}$ and $S_\text{B}$ with $k^2$ varying with $\phi$ and $T$. We search for at which temperature ($T_2$), the ion irradiation at $\phi=\phi_2$ reproduces the same $S_\text{V}$ or $S_\text{B}$ obtained from the neutron irradiation at $T=T_1$ and $\phi=\phi_1$ (e.g., Fig.\,\ref{fig:temperature_shift_Cr}).}  \\
\\
\end{tabular}
}
\end{ruledtabular}
\footnotetext[1]{Conditions for both neutron and ion irradiations. One can refer to Fig.\,\ref{fig:SB_Cr_k2} to help identifying the kinetic domains at different irradiation conditions and sink strengths.}
\end{table*}

{One objective of this work is to {provide quantitative temperature-shift criteria for ion-irradiation experiments aimed at} emulating RIS generated by neutron irradiation. 
{We ascribe the difference of structural evolution between neutron and ion irradiations to a difference of radiation flux. A change of temperature may compensate the effect of a change of the radiation flux on the vacancy profile or on the solute RIS. With this work, we can suggest temperature shifts that should be applied depending on the (evolving) microstructure and the RIS quantity that one wants to reproduce ($S_\text{V}$ or $S_\text{B}$).} 
Even though $S_\text{B}$ and $S_\text{V}$ are inter-dependent quantities (cf. Eq.\,\eqref{eq:SB_SV}), the behavior of solute RIS is very different from that of PDs, mainly because solute RIS results from a balance between the solute flux triggered by a PD driving force and the backward solute flux triggered by a solute concentration gradient, whereas such backward flux does not occur for PDs.
Another difficulty is that the behavior of both PDs and solutes depends not only on the radiation flux and temperature, but also on the evolving microstructure sink strength. The latter is a complex function of  temperature, radiation dose rate, and radiation dose (i.e., dose rate\,$\times$\,time), as shown in Fig. 18. 
{Besides, the evolution of $S_\text{V}$ and $S_\text{B}$ as a function of sink strength, radiation flux and temperature differs from one kinetic domain to another, and the extent of each kinetic domain in terms of temperature and radiation flux depends itself on the sink strength which evolves over time.}
Nevertheless, there are a few {limiting} cases (defined in Tab.\,\ref{tab:temperature-shift_case_definition}) which {provide some insights in this rather complex interplay and from which some quantitative temperature-shift criteria can be proposed}.

In case (i), the sink strength is assumed to be constant during irradiation. This is a good approximation for alloys with initially high dislocation density, for instance, cold-worked materials. At fixed sink strength, the amount of vacancy RIS, $S_\text{V}$, increases linearly with the ratio $\phi/D_\text{V}$ in the sink domain, and with $(\phi/D_\text{V})^{1/4}$ in the recombination domain. On the other hand, $S_\text{B}$ is independent of $\phi$ in the sink domain, whereas, in the recombination domain, it decreases with $\phi/D_\text{V}$. 
Thus, if the vacancy RIS is to be conserved from a neutron to a higher flux ion irradiation, we prescribe a shift of temperature such as to keep the ratio $\phi/D_\text{V}$ constant. {Concerning the RIS of PDs,} we recover the Mansur's invariant relation, which has been established in the recombination domain for swelling phenomena~\cite{Mansur1993}.
However, for the solute RIS in the sink domain, there is no need for a change of temperature to keep the  amount of solute RIS constant. 
{In the recombination domain, a temperature shift conserving the ratio $\phi/D_\text{V}$ does not necessary ensure a correct emulation of a neutron radiation-induced solute RIS. Therefore, one temperature shift only enables to reproduce one RIS quantity.} Indeed, the solute-PDs flux couplings leading to RIS are strongly non-linear and alloy specific functions of temperature. Nevertheless, we may use our temperature-flux maps to obtain an estimation of the temperature shift leading to the same amount of solute RIS. 
According to the maps of Fig.\,\ref{fig:SV_Fe-X} and Fig.\,\ref{fig:SB_Fe-X}, for $k^2=5\times10^{14}\,\text{m}^{-2}$, an emulation of neutron irradiation with a flux of $10^{-7}$ dpa/s at $T=360$\,K (i.e., $1000/T=2.75$) by means of an ion irradiation of $10^{-5}$\,dpa/s would require a shift of temperature $\Delta T\simeq+90\,$K for the PDs, and alloy dependent $\Delta T$ for solute RIS as listed in Table \ref{tab:temperature-shift_Fe-X}. Note that no temperature shift is proposed for Fe-Cr alloy because $S_\text{Cr}$ at $\phi=10^{-5}$\,dpa/s, at any temperature, is systematically smaller than that at $\phi=10^{-7}$ dpa/s and $T=360$\,K.
\begin{table}
\caption{\label{tab:temperature-shift_Fe-X}
Temperature shift required to simulate the solute RIS from the neutron irradiation with a flux of $10^{-7}$ dpa/s at 400\,K by means of an ion irradiation of $10^{-5}$\,dpa/s. The sink strength is assumed to be constant during irradiation {(case (i))}. 
}
\centering
\begin{ruledtabular}
{\begin{tabular}{c c c c c c c}
 & Fe-P & Fe-Mn & Fe-Cr & Fe-Si & Fe-Ni & Fe-Cu \\
\hline
$\Delta T$ [K] & $+95$ & $+40$ & --- & $+90$ & $+105$ & $+100$ \\
\end{tabular}
}
\end{ruledtabular}
\end{table}

In case (ii), both neutron and ion irradiations take place in the recombination domain, our results suggest that $S_\text{V}$ and $S_\text{B}$ are nearly independent of $k^2$ (cf. Figs.\,\ref{fig:SV_Fe-X_k2} and \ref{fig:SB_Fe-X_k2}). The temperature-shift criterion for $S_\text{V}$ is the same as the one in case (i). To estimate the temperature shift for $S_\text{B}$, we use the $\phi$--$T$ maps
of $S_\text{B}$ in the same way as presented in case (i).

In case (iii), both neutron and ion irradiations take place in the sink domain. We assume that the temperature shift is sufficiently small such that the variation of $\alpha_1$ can be neglected. Thus, $S_\text{B}$ only depends on the microstructure (cf. Eq.\,\eqref{eq:SB_elimination_dominant}). We assume that the time for the establishment of PD and solute RIS is much shorter than the characteristic time of the evolving microstructure. 
In this case, the temperature-shift criterion for $S_\text{B}$ is the one ensuring an invariant sink strength. Therefore, given the variations of $k^2$ with temperature and dose rate, the variations of $S_\text{B}$ should have the same trends. This is consistent with the experimental observation in Ref.\,\cite{Jiao2018}. In this experiment, the authors attempted to emulate the microstructure of a cold-worked 316-stainless steel produced by a neutron irradiation at 320\,\degree C by a self-ion irradiation at higher temperatures. In such cold-worked material, the sink density was relatively high; thereby the irradiation must take place in the sink domain. Authors in this study observed that self-ion irradiation at 380\,\degree C produces dislocation loop size and density which matched well with those obtained with neutron irradiation. In the same study, they showed that the RIS behaviors from these two irradiation conditions coincided as well. Therefore, this experiment shows that a relatively small temperature shift ($+60$\,\degree C in this experiment) ensuring an invariant microstructure (i.e., sink strength) is able to reproduce as well the RIS behaviors for materials irradiated in the sink domain. 

In case (iv), we assume that the evolution of the sink strength is not affected by the RIS of solutes. 
In this case, there are simulation methods and/or experimental studies yielding the evolution of the PDs microstructure with respect to the irradiation conditions and the radiation dose ~\cite{Soisson2016,Soisson2018}. 
Authors in Ref.\,\cite{Soisson2018} simulated the microstructural evolution of a Fe-Cr alloy irradiated by neutrons ($3.4\times10^{-7}\,$dpa/s) and ions ($5.2\times10^{-5}$\,dpa/s) at similar temperatures using cluster dynamics and atomic kinetic Monte Carlo simulations. Relying on their results, we can predict the evolution of the RIS behaviors. Note that their results indicate that the PD clusters are the major sinks. Due to the lack of information on cluster densities, we estimate the average distance between sinks directly from the sink strength by Eq.\,\eqref{eq:interplanr_sink_strength}. In Fig.\,\ref{fig:temperature_shift_Cr}, we plot $k^2$ (from Ref.\,\cite{Soisson2018}), $S_\text{V}$ and $S_\text{B}$ (from our calculation) as a function of the radiation dose. The evolution of $k^2$ indicates that, up to 0.01\,dpa for neutron irradiation and 0.1\,dpa for ion irradiation, the system is at the frontier between the recombination and sink domains. After these doses, the system is in the sink domain and the sink strengths of both neutron and ion irradiation conditions are close to each other. The calculated $S_\text{B}$ in the two irradiation conditions are as well very similar after 0.1\,dpa. This is because, in the sink domain, $S_\text{B}$ depends only on $\alpha_1$ and $k^2$ (as presented in case (iii)); since the temperatures are close in the two irradiation conditions, the calculated $S_\text{B}$ is nearly the same whenever the sink strengths are very close to each other. Below 0.01\,dpa, both $S_\text{V}$ and $S_\text{B}$ in the two irradiation conditions are different. Given the variation trends of the sink strength with the irradiation conditions, we propose a temperature-shift that would reproduce either the same $S_\text{V}$ or the same $S_\text{B}$ as in neutron irradiation from an ion irradiation experiment. As a qualitative approach, we assume that the sink strength $k^2$ is proportional to $S_\text{V}$. This approach should be reasonable because PD clusters are major sinks and their growth should be proportional to the PD segregation amount. {Thus, by assuming that the ion irradiation is in the recombination domain, we set $k^2$ as a linear function of $(\phi/D_\text{V})^{0.25}$. Hence, from the simulated $k^2(\phi)$ resulting from an ion irradiation~\cite{Soisson2018}, we can deduce the sink strength evolution at different temperatures. Relying on our model, we calculate the evolution of $S_\text{V}$ and $S_\text{B}$ from the ion irradiation at different temperatures. From these results, we find out at which temperature the evolution of $S_\text{V}$ or $S_\text{B}$ matches well with that obtained by neutron irradiation. By this approach, we obtain the temperature shifts of an ion irradiation ($5.2\times10^{-5}$\,dpa/s) required to emulate the RIS behaviors from neutron irradiation ($3.4\times10^{-7}$\,dpa/s) (cf. Fig.\,\ref{fig:temperature_shift_Cr}-(d)). For a dose below 0.01\,dpa, the temperature shift ($\Delta T$) required for an invariant $S_\text{V}$ is about +90$\degree$C and the one for an invariant $S_\text{B}$ is about +110$\degree$C. After 0.01\,dpa, $\Delta T$ for $S_\text{V}$ increases up to +200$\degree$C, whereas $\Delta T$ for $S_\text{B}$ notably decreases.} 

Apart from the simulation methods, direct observations of the microstructure may inform on the sink strength evolution. However, a precise estimation of the latter is difficult because small PD nano-clusters forming under irradiation are not detectable by current microscopy techniques. Nevertheless, investigating the variation of the solute RIS profiles with radiation flux and radiation dose should give an insight on the sink strength, provided the time scale of RIS is smaller than that of the microstructure evolution, so that we may assume steady-state solute RIS. For instance, we have shown that the RIS amount of solute atoms is directly related to the bulk concentration of vacancies. Therefore, measuring the solute RIS provides a way to estimate the bulk concentration of vacancies---thereby the global sink strength of the microstructure---provided the diffusion properties of PDs are known.

}

\begin{figure}[!htb]
    \centering
    \includegraphics[width=1.0\linewidth]{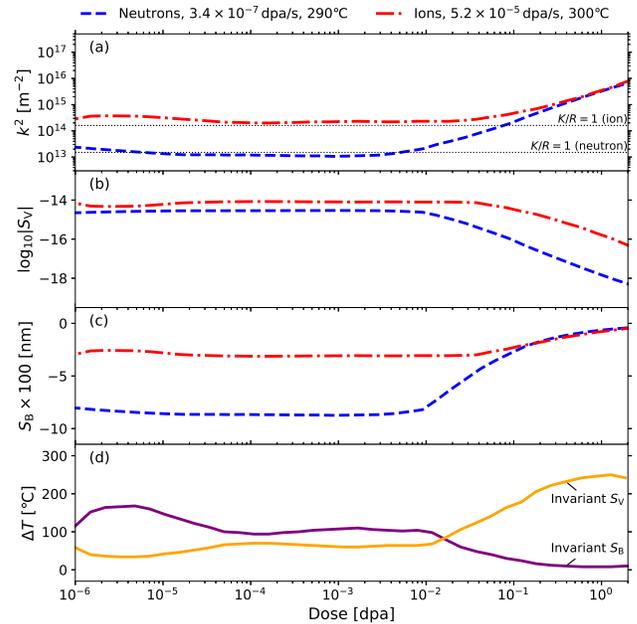}
    \caption{The evolution of (a) the sink strength $k^2$, (b) the amount of vacancy RIS $S_\text{V}$, and (c) the amount of solute RIS $S_\text{B}$ in the Fe-Cr alloy irradiated by neutrons and ions. The evolution of the temperature shift for ion irradiation that is required to emulate the neutron RIS is plotted in (d). The plots of $k^2$ are reproduced from the results in Ref.\,\cite{Soisson2018}. The dotted guiding lines obtained from $K/R=1$ are plotted in (a) to help identifying the kinetic domain. }
    \label{fig:temperature_shift_Cr} 
\end{figure}

\section{Conclusions \label{Sec:Conclusions}}


{
In this work, the cluster-expansion version of the self-consistent mean field theory is applied to calculate the transport coefficients of dilute iron-base alloys Fe-$B$ ($B$ = P, Mn, Cr, Si, Ni, and Cu) under irradiation. We add to the recent calculation of the transport coefficients~\cite{Messina2014,Messina2019} the contribution of forced atomic relocations (FARs)~\cite{Huang2019} 
From these transport coefficients, we compute the flux-coupling coefficients, the solute and vacancy diffusion coefficients, and the RIS factors with respect to temperature ($T$), radiation flux {($\phi$)}, and {point-defect} (PD) sink strength {($k^2$)}. We highlight the specificity of each alloy as well as the effect of FAR on these parameters.

We {provide} a general PD-RIS model yielding the concentration profile of vacancies in the vicinity of sinks in the three kinetic domains. The profile is divided into two regions: a region {of uniform vacancy concentration} far from the sinks where we account for PD production, recombination, and elimination at sinks, and a second region near the PD sinks where we neglect recombination reactions because PD concentrations are lower. This approximation leads to first-order differential equations that can be solved analytically. 

From the RIS factor relating the solute concentration gradient to the vacancy {one}, we deduce an analytical expression of the steady-state solute RIS profile. This analytical RIS model includes the full set of PD reactions, solute-PD interactions, and FAR mechanisms. 

We summarize below the most relevant results obtained from an application of the analytical results to {the investigated} dilute Fe-based alloys.
\begin{itemize}
    \item The consideration of the complete PD reactions enables a consistent investigation of RIS behaviors in all PD kinetic domains (recombination/sink/thermal). We show that the RIS kinetic domains are directly related to the PD kinetic domains, i.e., to the variation of PD concentration in the bulk. RIS profiles of PDs do not vary much with the chemical nature of the solute atom, whereas solute RIS profiles are very alloy-specific. In general, the RIS of PDs and solutes is favored in the sink domain because the rate of PD elimination at sinks is significant. In the recombination domain, even though the PD RIS amount is relatively small, the solute RIS amount can be high in certain alloys, such as for instance in Fe-Mn where the RIS factor $\alpha_1$ is relatively large.  
    \item The comparison between our results and a previous study~\cite{Martinez2018} highlights the sensitivity of RIS to recombination reactions. Models that would neglect these reactions would overestimate the vacancy concentration along the RIS profile, especially at low temperatures and sink strengths (i.e., in the recombination domain). 
    \item Parametric $T$--$\phi$--$k^2$ studies show that the effect of FAR on solute RIS is significant. At high sink strengths, FAR leads to a sharp decrease of solute RIS. Moreover, our results show that, among the investigated alloys, the effect of FAR is the most important in Fe-Ni and Fe-Cr systems.
    \item $T$--$\phi$--$k^2$ maps of the RIS amount of PDs and solute atoms can be used as a tool to provide quantitative temperature-shift criteria for the comparison between neutron and ion irradiation. We emphasize that these criteria are alloy and kinetic-domain specific. In the case where we may ignore the variation of sink strength with temperature and dose rate, for instance in alloys with a high sink density, we {show how to} rely on the maps to deduce the temperature shifts. Otherwise, in most cases, an estimation of the temperature shift requires {the knowledge of the} explicit relationship between the sink strength, temperature, and dose rate. 
\end{itemize}

Even though the present investigation is focused on dilute Fe-base binary alloys, the present RIS model can be applied to any alloys, provided that one is able to compute the RIS factor and the solute and PD diffusion coefficients. This RIS model can be extended to non-neutral PDs sinks by including the elastic interactions between PDs, solute atoms, and sinks into the calculation of the chemical potential gradients and the transport coefficients~\cite{PhysRevB.88.134108}.  

Finally, radiation-induced solute enrichment at sinks can exceed the alloy solubility limit and trigger the precipitation of a secondary phase. Such a radiation-induced precipitation phenomenon would require different boundary conditions on the solute RIS profile, as for example a backward solute diffusion set to zero. These points are left for future work. 
}

\appendix
\section{Mathematical descriptions} \label{sec:appendix_mathematical}
{
\subsection{Relation between \texorpdfstring{$C_\text{V}(z)$ and $C_\text{I}(z)$}{CV(z) and CI(z)}} \label{subsec:C_V and C_I}
Following Eq.\,\eqref{eq:CV_diffusion} and \eqref{eq:CI_diffusion}, we have:
\begin{align}
    0\,& = D_\text{V}\frac{\partial^2 C_\text{V}}{\partial z^2} - D_\text{I}\frac{\partial^2 C_\text{I}}{\partial z^2} \nonumber \\
       & = \frac{\partial^2 }{\partial z^2} \left( D_\text{V} C_\text{V} - D_\text{I} C_\text{I}\right).
\end{align}

Therefore, $D_\text{V} C_\text{V}(z) - D_\text{I} C_\text{I}(z) = K_2 z + K_3$, with $K_2$ and $K_3$ two integration constants to be determined. By symmetry, the PD flux at the mid-point ($z=0$) is zero, hence we have:
\begin{equation} \label{eq:C_PD_midpoint}
    \frac{\partial C_\text{V}}{\partial z}(z=0) = 0, \,\,\frac{\partial C_\text{I}}{\partial z}(z=0) = 0,
\end{equation}
and therefore $K_2=0$. Moreover, the PD concentrations at the sink are equal to the equilibrium concentrations. Therefore, we have:
\begin{equation} \label{eq:C_PD_sink}
    C_\text{V}(z=h/2) = C_\text{V}^\text{eq},\,\, C_\text{I}(z=h/2) = C_\text{I}^\text{eq}.
\end{equation}
Using Eq.\,\eqref{eq:C_PD_sink}, $K_3 = D_\text{V} C_\text{V}^\text{eq} - D_\text{I} C_\text{I}^\text{eq}$. Assuming that $ D_\text{I} C_\text{I}^\text{eq}\ll D_\text{V} C_\text{V}^\text{eq}$, we get $K_3\simeq D_\text{V} C_\text{V}^\text{eq}$. Accordingly, $C_\text{V}(z)$ and $C_\text{I}(z)$ are related by:
\begin{equation}
    D_\text{V} \left[C_\text{V}(z) - C_\text{V}^\text{eq}\right] = D_\text{I} C_\text{I}(z).
\end{equation}
}

\subsection{Introduction of the hypergeometric function \texorpdfstring{$_2F_1$}{2F1}} \label{subsec:hypergeometric_function}
The hypergeometric function $_2F_1(a,b,c,x)$ is defined by the series 
\begin{equation}
    _2F_1(a,b,c,x)=\sum_{n=0}^{\infty}\frac{(a)_n (b)_n}{(c)_n n!}x^n
\end{equation}
for $|x|\leq 1$, where $(a)_n$, $(b)_n$, and $(c)_n$ are the Pochhammer's symbol~\cite{Andrews1999} given by
\begin{equation}
    (a)_n = \left\{ \begin{array}{lr}
        a(a+1)\cdot(a+n-1), & \text{for } n\geq 1;\\
        1, & \text{for } n=0 \end{array}. \right.
\end{equation}

For $\text{Re}(c)>\text{Re}(b)>0$, we have
\begin{equation}
    _2F_1(a,b,c,x)=\frac{\Gamma(c)}{\Gamma(b)\Gamma(c-b)}f(a,b,c,x),
\end{equation}
where $\Gamma$ is the Gamma function~\cite{Andrews1999}, and
\begin{equation}
    f(a,b,c,x) = \int_0^1 t^{b-1}(1-t)^{c-b-1}(1-xt)^{-a}\text{d}t.
\end{equation}
We can deduce the integral (Eq.\,\eqref{eq:I}), $I$, from the hypergeometric function
\begin{align}
    I &= \left(\frac{h}{2}-l\right)\,b^{-2\alpha_1}\int_0^1\left[ 1-\left(\frac{h/2-l}{b}\right)^2 t \right]^{-\alpha_1} \frac{\text{d}t}{2\sqrt{t}} \nonumber\\
      &=\left(\frac{h}{2}-l\right)\,b^{-2\alpha_1}\,_2F_1\left(\alpha_1, \frac{1}{2};\frac{3}{2};\left(\frac{h/2-l}{b}\right)^2\right).
\end{align}

\bibliographystyle{apsrev4-2}
\bibliography{MyRef}

\end{document}